**Manganese oxide-functionalized graphene sponge electrodes for electrochemical chlorine-free disinfection of tap water**

*Anna Segues Codina[a,b], Natalia Sergienko[a,b], Carles M. Borrego[a,c], Jelena Radjenovic [a,d]**

[a] *Catalan Institute for Water Research (ICRA-CERCA), Scientific and Technological Park of the University of Girona, 17003 Girona, Spain*

[b] *University of Girona, Girona, Spain*

[c] *Group of Molecular Microbial Ecology, Institute of Aquatic Ecology, University of Girona, 17011 Girona, Spain*

[d] *Catalan Institution for Research and Advanced Studies (ICREA), Passeig Lluís Companys 23, 08010 Barcelona, Spain*

*\* Corresponding author:*

*Jelena Radjenovic, Catalan Institute for Water Research (ICRA), c/Emili Grahit, 101, 17003 Girona, Spain*

Phone: + 34 972 18 33 80; Fax: +34 972 18 32 48; E-mail: <u>jradjenovic@icra.cat</u>

**Abstract**

Low-cost reduced graphene oxide sponges functionalized with manganese oxide were used as electrodes for the disinfection of *Escherichia coli* in water. Manganese oxide was doped with amino groups ($Mn_xO_yNH_2$) to strengthen the bond with the graphene coating and improve the electrochemical stability of the sponge. The Mn II and Mn III incorporated into the graphene coating favored the formation of oxygen vacancies and enhanced the electric and catalytic properties of the anode. Electrooxidation of real tap water at 29 A m$^{-2}$ resulted in 2.2 log removal of *E. coli*. After storing the treated samples for 18 h at 25 and 37 °C, the removal of *E. coli* increased to 3.3 and 4.6 log, respectively, demonstrating the irreparable damage to the bacterial cells *via* low-voltage electroporation, and the key role of storage temperature in their further inactivation. The energy consumption of the system treating real low conductivity tap water was 1.2-1.6 kWh m$^{-3}$. Finally, experiments with intermittent current suggest the contribution of the pseudo-capacitance of the graphene sponge to electrochemical disinfection during the OFF periods. This study demonstrates the feasibility of manganese oxide-functionalized graphene sponges for the chlorine-free disinfection of water.

## 1. Introduction

The World Health Organization (WHO) estimates that 785 million people do not have access to a reliable source of drinking water and that 829.000 persons die annually from diarrheal illnesses caused by the lack of clean water, hygiene, and sanitation. The situation is expected to worsen in the upcoming years due to a decrease in the availability of clean water caused by the population growth, climate change and water pollution. Around 50% of the Earth's inhabitants will live in regions of water scarcity by 2025 [1]. Therefore, water treatment and disinfection will become of vital importance to ensure access to safe water, especially in low-income and rural areas where this problem is intensified.

Chlorination is widely used as a disinfection treatment worldwide, however, it leads to the formation of toxic disinfection by-products (DBPs) [2]. UV-based technologies have a high energy consumption, and the photochemical damage caused to microorganisms is often reversible, *i.e.*, they can be reactivated after the treatment [3]. Ozonation has a high cost of operation and can also generate toxic DBPs [4]. Furthermore, these methods are not easy to apply in rural or developing regions, for economic and maintenance reasons.

Electrochemical processes have emerged as a promising technology for water treatment and disinfection and are particularly well-suited for decentralized and distributed systems for several reasons, namely: *i)* they do not require any addition of chemicals; *ii)* they have a small footprint with modular treatment units; *iii)* they can be easily powered by renewable energy; and *iv)* they operate at ambient pressure and temperature [5,6]. Boron-doped diamond (BBD) [7,8], Magnéli phase $Ti_4O_7$ [9] and mixed metal oxide (MMO) electrodes [10,11] are examples of widely studied anodes for electrochemical water treatment and disinfection. However, high cost of these electrodes and/or formation of toxic chlorinated by-products due to the electro-oxidation of

chloride —a naturally occurring anion— limits their applicability for electrochemical point-of-use and point-of-entry disinfection of water [12]. Besides organochlorine byproducts, when using electrodes with high oxidizing power such as BBD and $Ti_4O_7$ anodes, chlorate ($ClO_3^-$) and perchlorate ($ClO_4^-$) are formed at current densities typically applied in electrochemical water treatment (*e.g.*, ≥100 A m$^{-2}$) [13,14]. Recently developed graphene sponge electrodes overcome this major bottleneck of existing anodes, as they do not form any chlorate and perchlorate, even in the presence of high chloride concentrations, and have a very low electrocatalytic activity towards chloride electro-oxidation to free chlorine (current efficiency <0.1 %) [15]. Furthermore, low-cost bottom-up synthesis method allows easy incorporation of dopants into the reduced graphene oxide (RGO) coating and thus tailoring its electrocatalytic activity. Graphene sponge electrodes doped with atomic nitrogen and boron were previously employed for the inactivation of *Escherichia coli* (*E. coli* ATCC 700078), whereas the pseudo-capacitance of graphene enabled the operation of the reactor using intermittent current [16].

In this study, we developed graphene sponge electrodes functionalized with manganese oxides and investigated their performance for chlorine-free electrochemical disinfection of a more resistant strain of *E. coli* (ATCC 25922). Manganese oxide is a low-cost and earth-abundant catalyst with a high toxicity threshold, which makes it very well-suited for water treatment applications [17]. It can exist in several oxidation states, has a high catalytic activity, and has been widely investigated for the removal of heavy metals [18], and catalytic [19] and electrocatalytic oxidation of organic pollutants [20]. Furthermore, due to its high capacitance, manganese oxide-based electrodes have been used in energy storage applications such as supercapacitors and batteries [21].

To exploit the capacitance and oxidative properties of manganese oxide for electrochemical disinfection, manganese oxide ($Mn_xO_y$)-functionalized graphene sponge electrodes were employed as anodes for electrochemical disinfection of tap water. Doping of the $Mn_xO_y$ with amino groups was used to enhance the electrocatalytic stability of the electrode and prevent the release of manganese during anodic polarization, forming covalent bonds (Mn-N-O) that strengthen the connection between manganese oxide and the RGO coating [22]. $NH_2$ doping of metal or metal oxides linked to C-based materials has been used to improve the structure and the performance of the material for their use as catalysts in the oxidation of organic compounds or photocatalytic reduction of $CO_2$ (**Table S1**). Disinfection experiments were performed in a one-pass flow-through reactor at different current densities and using real tap water. The inactivation of *E. coli* was verified by the additional overnight storage experiments at two temperatures (*i.e.*, 25 and 37ºC), and scanning electron microscopy (SEM) was employed to assess the morphology of *E. coli* cells before and after the treatment. Experiments applying intermittent current were performed to exploit the pseudo-capacitance of the material and lower the energy consumption. This study demonstrates that manganese oxide-functionalized graphene sponge electrodes can inactivate resistant *E. coli* in real tap water at very low applied anodic currents, and without the generation of chlorine, chlorate and perchlorate.

## 2. Materials and methods

### 2.1. Chemicals and reagents

Graphene oxide (GO) water dispersion (4 g L$^{-1}$) was purchased from Graphenea and the mineral wool from Diaterm. Potassium permanganate, urea, 3-aminopropyltriethoxysilane (APTES), Chromocult Coliform Agar and phosphate salts were provided by Sigma Aldrich. Manganese sulphate monohydrate was purchased from Honeywell and toluene from PanReac AppliChem.

Ringer and medium LB were provided by Scharlau. The *E. coli* strain ATCC 25922 was provided by LGC standards.

## 2.2. Synthesis of the graphene sponges

Graphene-based sponges were prepared using a modified bottom-up hydrothermal synthesis method, previously developed by Baptista-Pires et al [15] (**Figure S1**). Nitrogen-doped reduced graphene oxide (NRGO) sponge, which was employed as a cathode in the flow-through reactors, was synthesized using urea as an N-source [15,16]. Undoped graphene sponge was also synthesized using the same procedure, to allow the comparison of manganese oxide-based materials with the unmodified RGO electrode.

Four different $Mn_xO_y$-functionalized graphene sponges were synthesized, to investigate the impact of the catalyst loading and manganese oxide modifications on the anode performance and stability. To produce the starting $GO/Mn_xO_y$ dispersions for the synthesis of $Mn_xO_y$-functionalized graphene sponges at low ($LMn_xO_yRGO$) and high $Mn_xO_y$ concentration ($HMn_xO_yRGO$), 10 mL of $MnSO_4$ of varying concentrations (*i.e.*, 0.09 and 0.9 M) were added dropwise to 10 mL of $KMnO_4$ (0.06 and 0.6 M, respectively), stirred for 15 min, mixed with 50 mL of 4 g L$^{-1}$ GO dispersion, and sonicated for 30 min.

To improve the electrochemical stability of the $Mn_xO_y$-functionalized graphene sponge anode, amino-functionalized manganese oxide ($Mn_xO_y$-$NH_2$) was synthesized according to the method proposed by Xue et al [22]. First, $Mn_xO_y$ was synthesized by dissolving $MnSO_4 \cdot H_2O$ (1.90 g) and $KMnO_4$ (1.76 g) in 70 mL of MiliQ water. The solution was stirred for 45 minutes and placed in an autoclave reactor for 12 h at 140 °C. The obtained solid was filtered, washed with MiliQ water and ethanol, and dried in the oven overnight at 60 °C. To functionalize the $Mn_xO_y$, the dried solid was added to 130 mL of toluene at 80 °C and stirred for 30 minutes, followed by the addition of 13 mL of 3-aminopropyltriethoxysilane (APTES) and stirring for 8 h at 80 °C. The resulting $Mn_xO_y$-$NH_2$ solid was filtered, washed with MiliQ water and ethanol, and dried in the oven overnight at 60 °C. To produce the starting GO/$Mn_xO_y$-$NH_2$ dispersions for graphene sponges with low (L$Mn_xO_yNH_2$RGO) and high (H$Mn_xO_yNH_2$RGO) $Mn_xO_y$-$NH_2$ concentration, 1 and 2 g of $Mn_xO_y$-$NH_2$, respectively, was added to 50 mL of GO (4 g $L^{-1}$) and 20 mL of MiliQ water, and ultrasonicated for 1 hour.

## 2.3. Characterization of the material

The morphology of the sponges was studied by scanning electron microscopy (SEM), using a JSM-7001F field emission scanning electron microscope (FESEM, JEOL, Japan). The elemental mapping images were performed by coupling the SEM to an energy-dispersive X-ray spectrometer (EDS) (X-Max 20 mm2, Oxford Instruments, UK). The chemical composition of the surface of the sponges and the $Mn_xO_y$-based compounds was done by X-ray photoelectron spectroscopy (XPS), using a PHOIBOS 150 electron analyzer (Specs, Germany) with a monochromatic aluminium Kalpha X-ray source. The spectra were calibrated at C1s 284.6 eV. Electrochemical characterization of graphene sponge electrodes was performed using cyclic voltammetry (CV) and electrochemical impedance spectroscopy (EIS) analyses. The EIS curves were fitted with the BioLogic EC-Lab software and the values were used to calculate the ohmic drop-corrected anode potentials and the double-layer capacitance ($C_{dl}$).

## 2.4. Electrochemical disinfection experiments

Manganese oxide-functionalized graphene sponge anode and N-doped graphene sponge cathode, each with a projected surface area of 17.34 $cm^2$ and a thickness of 1 cm, were connected to stainless steel current feeders, separated by a thin polypropylene mesh to avoid short-circuiting and placed into a cylindrical flow-through reactor (**Figure S2**). All experiments were performed in one-pass mode at a flow rate of 5 mL min$^{-1}$ using a digital gear pump (Cole-Parmer), which corresponds to a hydraulic residence time (HRT) of 3.5 min and an effluent flux (LMH) of 175 L m$^{-2}$ h$^{-1}$. Three different supporting electrolytes were employed, phosphate buffer (10 mM, pH 7, 1 mS cm$^{-1}$), phosphate buffer (5 mM, 0.47 mS cm$^{-1}$), and tap water (0.42 mS cm$^{-1}$) (**Table S2**), spiked with $10^7$ CFU mL$^{-1}$ of *E. coli* ATCC 25922, a commonly used indicator bacterium for electrochemical [9,23] and chlorine-based disinfection [24]. Prior to the experiments, all residual chlorine in the employed tap water was removed by intense overnight stirring, to exclude its participation in the *E. coli* inactivation. The concentration of free and total chlorine in the stirred water was below the detection limit (<0.05 mg L$^{-1}$). Chronopotentiometric experiments were conducted at different current densities, i.e., 58, 115 and 173 A m$^{-2}$ for 10 mM phosphate buffer and 14, 20 and 29 A m$^{-2}$ for tap water, due to the differences in the electrical conductivities of the electrolytes. Current/potential control was achieved using a leak-free Ag/AgCl reference electrode (Harvard Apparatus) placed between the cathode and the anode and a multi-channel potentiostat/galvanostat VMP-300 (Biologic). The flow-through reactor was first operated in the initial open circuit (OC$_i$) to verify the loss of the bacteria in the absence of current, followed by the chronopotentiometric experiments, and a final OC run (OC$_f$) after the current was switched off, to verify if there was any electrosorption and posterior release of *E. coli* cells without their inactivation. To exploit the pseudo-capacitance of the material, experiments with the application of intermittent current were performed using HMn$_x$O$_y$NH$_2$RGO anode in tap water and applying 29 A m$^{-2}$ in 75s ON- 30s OFF cycles.

**2.5. Microbiological analyses**

*Preparation of stock cultures and determination of the cell concentration*: *E. coli* (ATCC 25922) was grown overnight at 37 ºC in Luria-Bertani broth. Overnight cultures (100 mL) were centrifuged at 3,500 rpm for 10 min in a benchtop centrifuge (Eppendorf 5804R). Supernatants were carefully discarded and pelleted cells were resuspended in phosphate buffer or tap water. Before starting each experiment, an aliquot of the corresponding suspension was serially diluted in a sterile Ringer solution, and decimal dilutions were filtered through sterile 0.22 µm pore, 47 mm diameter, cellulose esters membranes (Millipore) that were then placed onto Chromocult Coliform agar plates. Plates were incubated overnight at 37 ºC and dark blue colonies corresponding to *E. coli* were quantified. The average concentration of *E. coli* in the working stocks was $5\times10^7 \pm 1\times10^7$ CFU mL$^{-1}$.

During each experiment, samples were collected at different time intervals and stored at 4 °C until plating. Collected samples were serially diluted in sterile Ringer solution and then filtered as described above. Dark blue colonies were counted, and the concentration of *E. coli* (in CFU mL$^{-1}$) was calculated before and immediately after each electrochemical treatment and reactivation experiments (see below). No colonies other than presumptive *E. coli* were found in any of the treatments (data not shown).

*Reactivation experiments*: We assessed the potential capacity of *E. coli* cells to reactivate from the electrochemical damage and regrowth. Experiments were conducted by incubating the treated samples overnight (~18 h) at 25 ºC (room temperature) and 37 ºC (the optimal growth temperature for *E. coli*). After incubation, samples were diluted and plated as described above.

*Cell morphology*: To assess the effect of the electrochemical treatment on the morphology of *E. coli* cells, treated samples were fixed and visualized under FESEM, as previously described [16]. Samples were fixed with 1 mL of glutaraldehyde 2.5% in sodium cacodylate (0.1 M, pH 7.4) and stored for 2 h at 4 °C. Afterwards, the samples were washed with sodium cacodylate (0.1 M) and dehydrated by increasing the concentration of added ethanol (from 50% to absolute ethanol) every 20 min. Once dehydrated, the critical point method (K850 critical point dryer, Emitech) was used to dry the samples. The samples were coated with carbon (K950 Turbo Evaporator, Emitech) to create a conductive layer. Observations were carried out in a FESEM (S4100, HITACHI, Japan). Images were digitally recorded and processed with the Quartz PCI program (Quartz Imaging Corporation, Canada) with a resolution of 2516 × 1937 pixels. All experiments were performed in duplicate, and the results are presented as mean values with their standard deviations (SDs).

## 2.6. Analytical methods

Steady state concentrations of the electrochemically produced oxidants were measured at the highest applied current of 29 A m$^{-2}$ in tap water. Ozone and free chlorine were measured with the Chlorine/Ozone/Chlorine dioxide cuvette tests LCK 310 (Hach Lange). Hydrogen peroxide ($H_2O_2$) was determined with a spectrophotometric method with a low quantification limit (1 µM), in which $H_2O_2$ reacts with the added copper(II) and 2,9-dimethyl-1,10-phenanthroline (DMP) to form a Cu(DMP)$_2^{+}$ complex, which is detected at 454 nm [25]. Steady state concentration of the electro-generated hydroxyl radicals ($\cdot$OH) was evaluated using terephthalic acid (TA) as a probe compound (**Figure S3**), whereas the TA concentration was measured spectrophotometrically at 240 nm (quantification limit of 1.98 mg L$^{-1}$) [26]. To assess the stability of the manganese oxide-functionalized graphene sponge anodes, effluent samples of the reactors operating at 20 and 29 A m$^{-2}$ with tap were analyzed for the total manganese using inductively coupled plasma-optical emission spectroscopy (ICP-OES) (Agilent 5100).

## 3. Results and discussion

### 3.1. Characterization of the graphene-based sponges and the Mn$_x$O$_y$-based compounds

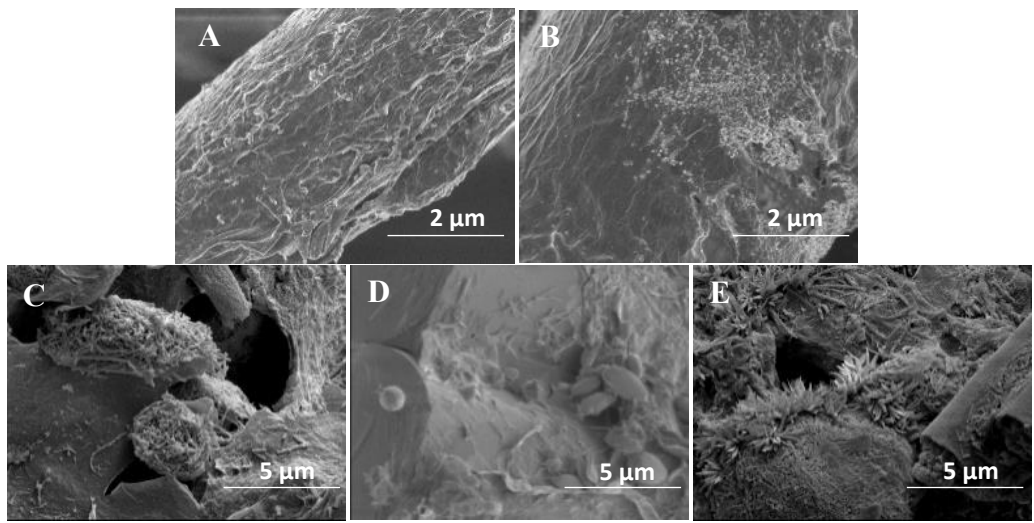

**Figure 1.** Scanning electron microscopy (SEM) images of the surface of **(A)** RGO, **(B)** LMn$_x$O$_y$RGO, **(C)** LMn$_x$O$_y$NH$_2$RGO, **(D)** HMn$_x$O$_y$RGO and **(E)** HMn$_x$O$_y$NH$_2$RGO.

The morphology of the surface of the sponges was studied by SEM (**Figure 1**). The RGO images show that the hydrothermal treatment creates a wrinkled graphene sheet coating on the mineral wool. The Mn$_x$O$_y$-based compounds can be observed on the surface of the sheets in various morphologies. Small particles were seen in LMn$_x$O$_y$RGO, which turn into nanorods HMn$_x$O$_y$RGO as the concentration of Mn$_x$O$_y$ is increased. The formation of the manganese oxide consisted of an initial nucleation phase and the growth of the crystal afterwards [27]. Higher concentrations of Mn$_x$O$_y$ led to larger nuclei and crystals of the resultant metal oxide [28]. LMn$_x$O$_y$NH$_2$RGO and HMn$_x$O$_y$NH$_2$RGO have needle-shaped particles. The Mn$_x$O$_y$ before the amino doping received a hydrothermal treatment, without GO, that was not applied to the undoped Mn$_x$O$_y$ of LMn$_x$O$_y$RGO and HMn$_x$O$_y$RGO. Previous study [29] reported that an increase in GO concentration inhibits the growth of manganese oxide into needles keeping the particles separated. Hence, the absence of GO in the additional hydrothermal treatment of the $Mn_xO_y$-$NH_2$ allows the growth of the manganese oxide particles as needles. Similar morphology was reported by Xu et al [22] in the $Mn_xO_y$ before and after the amino doping. The elemental mappings (**Figure S4**) show the three main parts of the material: Si, O, Na and Ca of the mineral wool, C and O of the RGO coating and, Mn and O of the $Mn_xO_y$. The elements of the mineral wool, especially silicon, showed the most intense signal as the RGO coating was thin compared to the size of the mineral wool template. The Mn was observed on the needle-shaped or nanorod particles, confirming the presence of the manganese-based compounds at the sponge's surface.

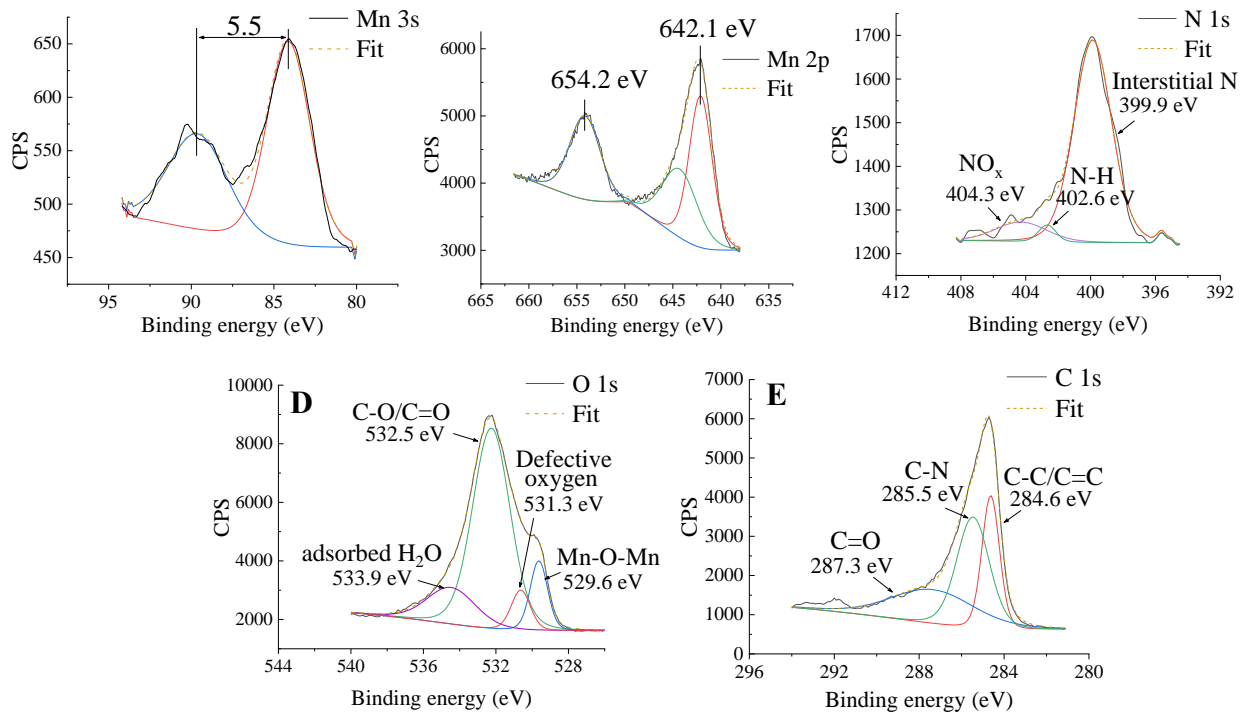

**Figure 2.** X-ray photoelectron spectroscopy (XPS) spectra of $HMn_xO_yNH_2RGO$ anode for **(A)** Mn 3s, **(B)** Mn 2p, **(C)** N 1s, **(D)** O 1s, and **(E)** C 1s.

The chemical composition of the surface of the sponges was characterized by the XPS analysis (**Figure 2**). The oxidation state of the $Mn_xO_y$ was determined with the peak splitting of the Mn 3s doublets, as it is considered more reliable than the use of the location of the Mn 2p peaks [30] (**Figure S5**). The values of the Mn 3s separation correlated to the oxides and valences of Mn reported in the literature, being 4.5 eV for $MnO_2$ (Mn IV), 5.2 eV for $Mn_2O_3$ (Mn III), 5.4 eV for $Mn_3O_4$ (Mn II/Mn III) and 5.8 eV for MnO (Mn II) [30,31] (**Figure S6**). As shown in **Table S3**, the XPS analyses of the $Mn_xO_y$ and $Mn_xO_y$-$NH_2$ indicated the presence of Mn in the oxidation state of Mn IV, thus implying that the amino-doping did not modify the Mn valence state. $LMn_xO_yRGO$ and $LMn_xO_yNH_2RGO$ have 6 and 11.1 % of Mn, respectively (**Table S4**). These low atomic concentrations resulted in a non-reliable determination of Mn valence state from the Mn 3s spectra. $HMn_xO_yRGO$ and $HMn_xO_yNH_2RGO$ have higher concentrations of Mn, 28.6 and 17.9% respectively, with a predominant presence of Mn II and Mn III. The interaction between RGO and the $Mn_xO_y$-based compounds decreased the oxidation state of the metal. This decrease in the oxidation state favored the formation of oxygen vacancies, which are reported to enhance the electrical conductivity, act as active sites for reactions and increase the material capacitance [32].

The O 1s spectrum of all the samples (**Figure S7**) can be deconvoluted into Mn-O-Mn (529.5-530.5 eV) [33], defective oxygen (530.7-531.6 eV) [34] and adsorbed water at the highest binding energy [35]. A peak corresponding to C=O/C-O appeared at 532.5-533.8 eV [36] in the sponge samples due to the RGO coating, and in the $Mn_xO_y$-$NH_2$ due to the process of amino doping which was performed with toluene and APTES, both containing carbon that could have reacted or been adsorbed onto the resulting $Mn_xO_y$-$NH_2$. $Mn_xO_y$ and $Mn_xO_y$-$NH_2$ had similar amounts of oxygen defects (31.3-38.1%) (**Table S5**). Also, $LMn_xO_yRGO$ and $LMn_xO_yNH_2RGO$ sponges had 21 and 23.3% of defective oxygen, respectively, whereas $HMn_xO_yRGO$ sponge showed higher quantities of oxygen defects (52.4%) than $HMn_xO_yNH_2RGO$ (36.7%).

The C 1s spectra of the sponges (**Figure S8**) showed three constituents that correspond to the RGO coating: C=O at 287.3-288.4 eV, C-N at 285.1-285.8 eV and C-C/C=C at 284.4-284.6 eV [37]. Also, $Mn_xO_y$-$NH_2$ showed the peaks of C-N, C-O and C=O. The content of N in the $Mn_xO_y$-doped sponges is 0.5-0.7% and mainly comes from the commercial GO solution used in the graphene sponge synthesis (**Table S4**). The synthesized $Mn_xO_y$-$NH_2$ contained 5.1% of N, identified as interstitial N (Mn-O-N or Mn-N-O) (400 eV) [38] and N-H (402 eV) [39] (**Figure S9**), confirming that the manganese oxide was functionalized. $LMn_xO_yNH_2RGO$ and $HMn_xO_yNH_2RGO$, contained 2.9 and 2.8% of N, respectively, and showed the incorporation of the amino-doped compound as three peaks were observed: interstitial N (399.8-399.9 eV), N-H (402-402.7 eV) and $NO_x$ (404.4-404.7 eV) [38]. The main peak in the $Mn_xO_y$-$NH_2$ and the resultant sponges was the interstitial N, which corresponded to the covalent bond formed in the doping and functionalization processes (Mn-O-N or Mn-N-O). Some samples showed a peak of $NO_x$ due to the ability of manganese oxide to adsorb nitrogen oxide groups [40].

**Figure S10** shows the Nyquist plots obtained by the EIS analyses in tap water for the four $Mn_xO_y$ and $Mn_xO_y$-$NH_2$ functionalized graphene sponge anodes. The $R_{ct}$ values are similar for the four anodes (from 25.8 to 27.4 Ω). Manganese oxides have poor electrical conductivity [41], thus a higher concentration of the metal oxide would increase the resistance of the system. In this case, the bond of $Mn_xO_y$ with the RGO and the presence of oxygen vacancies enhance the electron transport, ensuring an electrically conductive material even with a high concentration of $Mn_xO_y$ present in the graphene coating. The Warburg line is the straight line over the Nyquist semicircle and characterizes the ion diffusion from the electrolyte into the electrode material. $LMn_xO_yNH_2RGO$ and $HMn_xO_yRGO$ have a more vertical line, which indicates an enhanced ion diffusion. The double layer capacitance of the system was calculated from the EIS measurement to be 0.07 F g$^{-1}$ for RGO, 0.08 F g$^{-1}$ for the low-concentrated anodes and 0.11 F g$^{-1}$ for the high-concentrated. Thus, the capacitance of the graphene sponges increased with the functionalization and loading of RGO with manganese oxide.

**3.2. Electrochemical disinfection in phosphate buffer**

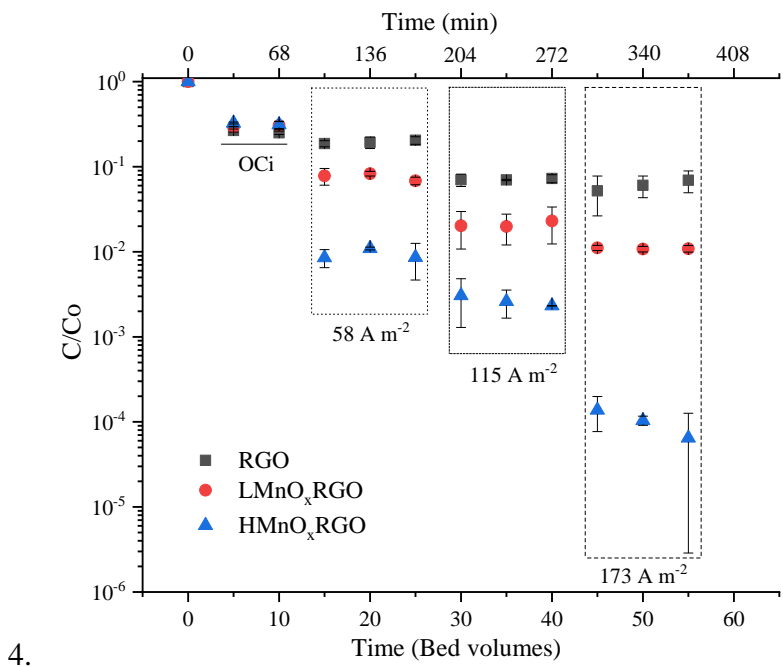

4.

**Figure 3.** Electrochemical inactivation of *E. coli* ATCC 25922 in the initial open circuit (OC$_i$) and at 58, 115 and 173 A m$^{-2}$ using 10 mM phosphate buffer (pH 7) as supporting electrolyte, and RGO, LMn$_x$O$_y$RGO and HMn$_x$O$_y$RGO anode, all versus the NRGO cathode.

Electrochemical inactivation of *E. coli* was first investigated in a low-conductivity electrolyte (10 mM phosphate buffer). **Figure 3** shows the removal of *E. coli* when using 10 mM phosphate buffer as supporting electrolyte, and three different graphene sponge anodes —RGO, LMn$_x$O$_y$RGO and HMn$_x$O$_y$RGO anode— versus the NRGO cathode. Samples were taken every 34 minutes, which correspond to 5 bed volumes, to make sure the reactor has reached the steady state. Regardless of the anode functionalization, around 0.5 log of *E. coli* were removed in the initial open circuit (OCi) due to the antimicrobial properties of RGO [42]. Graphene-based coatings are considered to inactivate the bacteria *via* membrane stress, caused by the direct contact between the cell and the sharp graphene nanosheets that damage the membrane and cause cell death [42]. Some studies suggest that graphene-based compounds can also inactivate the bacteria by reactive oxygen species-independent oxidative stress, which consists of the disruption of the cell functions via the oxidation of essential constituents, without the participation of oxidative species [43]. The *E. coli* removal was enhanced at all three anodes with the application of current. At the lowest applied current density of 58 A m$^{-2}$, the RGO anode achieved 0.7 log removal of *E. coli*, whereas LMn$_x$O$_y$RGO and HMn$_x$O$_y$RGO anodes achieved 1.1 and 2 log removals, respectively. Thus, the functionalization of the graphene sponges with Mn$_x$O$_y$ had a positive effect on the electrochemical disinfection, which was enhanced at a higher concentration of manganese oxide. The impact of manganese oxide functionalization was more obvious at higher current density, and at 173 A m$^{-2}$ the observed *E. coli* removals were 1.2, 2 and 4 log removal for RGO, LMn$_x$O$_y$RGO and HMn$_x$O$_y$RGO anode, respectively. The ohmic drop corrected potentials at 173 A m$^{-2}$ ranged from 2.3 to 4.8 V/SHE (**Table S6**). Nevertheless, prolonged reactor operation revealed insufficient electrocatalytic stability of the synthesized LMn$_x$O$_y$RGO and HMn$_x$O$_y$RGO electrodes, as evidenced by the formation of Mn phosphate precipitate; this was exacerbated for graphene sponge anode functionalized with the higher concentration of Mn$_x$O$_y$. The release of Mn is associated to the disproportionation of Mn III to Mn$^{2+}$ and MnO$_2$ [44] and, the interaction between Mn and phosphate, which have a high affinity that leads to the formation of a manganese phosphate precipitate. Although phosphate buffer is a well-suited supporting electrolyte to investigate the electrochemical water treatment at circumneutral pH, high concentration of phosphate leads to the dissolution of the manganese oxide incorporated into the graphene coating and compromises the stability of the $Mn_xO_y$-functionalized graphene sponge anode. Whereas 10 mM phosphate buffer contains ~949 mg L$^{-1}$ of phosphate, in the case of tap water and most natural waters used as drinking water supply, the concentration of phosphate is below 0.05 mg L$^{-1}$ [45]. Thus, to avoid the detrimental impact of the highly concentrated phosphate solution, further electrochemical disinfection experiments were conducted using real tap water, which had less than 0.008 mg L$^{-1}$ of phosphate (**Table S2**).

### 3.3. Electrochemical disinfection in tap water

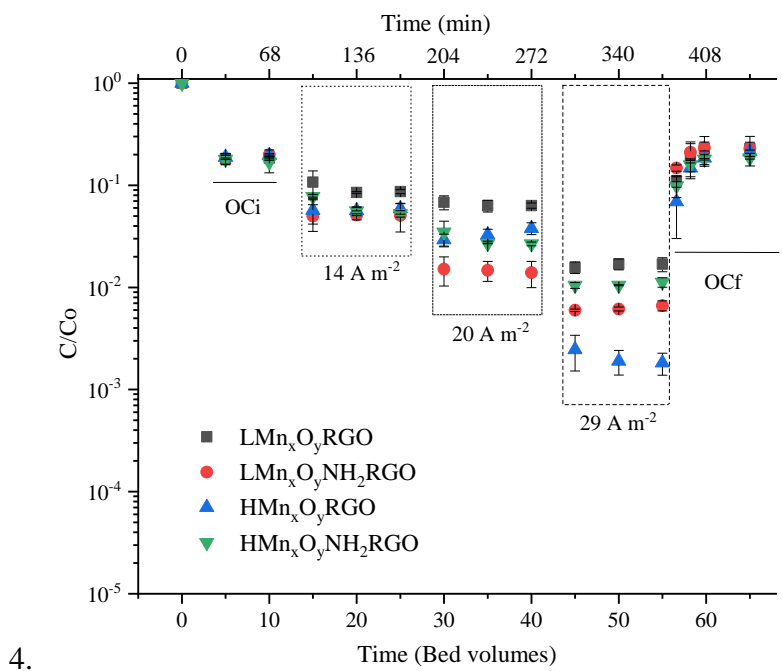

4.

**Figure 4**. Electrochemical inactivation of *E. coli* ATCC 25922 in tap water, in the initial open circuit (OC$_i$), at 14, 20 and 29 A m$^{-2}$ and in the final OC (OC$_f$), using LMn$_x$O$_y$RGO, LMn$_x$O$_y$NH$_2$RGO, HMn$_x$O$_y$RGO and HMn$_x$O$_y$NH$_2$RGO anode, all versus NRGO cathode.

Doping of the Mn$_x$O$_y$ with amino groups was employed to strengthen the bond between the Mn and the RGO, minimize the Mn$^{2+}$ release and thus enhance the stability of the manganese oxide-functionalized graphene sponge anode [22]. **Figure 4** shows the achieved removals of *E. coli* using $Mn_xO_y$ and $Mn_xO_yNH_2$-functionalized graphene sponge anodes in tap water. In the $OC_i$, 0.7 logs of *E. coli* were removed in all four systems, and this removal was progressively enhanced with the application and increase of the anodic current density. For example, for $HMn_xO_yRGO$ anode, the removal was increased from 1.2 to 1.5 and 2.7 log when increasing the current from 14 to 20 and 29 A m$^{-2}$. Although these values are lower compared with the ones achieved in the 10 mM phosphate buffer, this is likely due to somewhat lower ohmic drop-corrected anode potentials and potential gradient in tap water, and thus less effective inactivation of the bacterial cells via electroporation, as discussed below (**Table S7**). For example, the ohmic drop-corrected potential of $HMn_xO_yRGO$ anode was 2.7 V/SHE for 29 A m$^{-2}$ in tap water and 4.3 V/SHE for 58 A m$^{-2}$ in phosphate buffer (**Table S6**). Lower applied current densities in tap water were conditioned by its very low electrical conductivity (0.4 mS cm$^{-1}$) compared with the buffer solution (1.1. mS cm$^{-1}$). Given that the employed tap water contained ~36 mg L$^{-1}$ of chloride, the concentration of the electro-generated chlorine was measured in tap water at the highest applied current density (29 A m$^{-2}$), and the values were below the detection limit (<0.05 mg L$^{-1}$). This result excluded the possibility that the inactivation of *E. coli* was caused by the residual free chlorine and confirmed the low electrocatalytic activity of the graphene-based sponges towards chlorine production.

The lowest removal of *E. coli* was obtained for $LMn_xO_yRGO$ anode, with 1.8 log removal at 29 A m$^{-2}$, which was increased to 2.7 log removal at the $HMn_xO_yRGO$ anode. $HMn_xO_yRGO$ has a higher ohmic drop-corrected potential than $LMn_xO_yRGO$ (2.71 and 1.83 V, respectively) due to the higher amount of manganese oxide having semiconductor properties. $LMn_xO_yRGO$ has higher charge-transfer resistance ($R_{CT}$) and a less vertical Warburg line, indicating a more difficult ion diffusion (**Figure S10**). The electrochemically produced oxidants have been reported to contribute to the inactivation of bacteria [9], however, the concentrations of $O_3$ (1.2-2.9 µM), $H_2O_2$ (15-22 µM) and

·OH radicals (3-4 x $10^{-8}$ µM) produced were similar in all four investigated systems (**Figure S11**). Even though the measurements of oxidant species were done immediately after sampling the effluent, the values reported are conservative and residual as the exact number is unknown due to possible reactions inside the reactor (*e.g.*, $H_2O_2$ decomposition to ·OH at the N-RGO cathode). The higher resulting anode potential of the $HMn_xO_y$RGO anode, in addition to the presence of more oxygen vacancies and catalytically active sites for the inactivation of the bacteria, cause more damage to the bacterial cells by electroporation and thus result in a higher log removal of *E. coli* (see Section 3.6).

The impact of the manganese oxide concentration was not as evident for the $Mn_xO_yNH_2$-functionalized anodes, with the $LMn_xO_yNH_2$RGO anode yielding similar inactivation of *E. coli* to the $HMn_xO_yNH_2$RGO anode (2.2 and 2 log removal, respectively). This can be attributed to similar ohmic drop-corrected anode potentials for the low and high $Mn_xO_yNH_2$ concentration (1.67 V and 1.70 V, respectively), leading to similar cell damage. $LMn_xO_yNH_2$RGO and $HMn_xO_yNH_2$RGO anode have similar charge transfer resistances ($R_{CT}$=26.6 and 25.9 Ω, respectively), however, the Warburg line is more vertical for $LMn_xO_yNH_2$RGO, indicating an enhanced ion diffusion (**Figure S10**). Thus, enhanced migration of the negatively charged *E. coli* cells towards the anode surface is likely responsible for somewhat enhanced disinfection performance of the graphene sponge anode with the lower concentration of $Mn_xO_yNH_2$.

When the current was switched off, the concentration of the *E. coli* gradually returned to its initial values in the $OC_f$, demonstrating that the removal observed was caused by the direct effect of the applied current on the cell viability, and not only by the electro-sorption of *E. coli* at the graphene sponge electrode surface. The quantity of the bacteria measured in the effluent in $OC_f$ did not reach the influent concentration of *E. coli* until around 34 minutes (5 bed volumes) have passed, likely due to the pseudo-capacitance of the graphene sponge anodes and retention of charge after the current was switched off.

**Table 1.** Concentration of total dissolved manganese in the outlet solution at 20 and 29 A m$^{-2}$ applied current densities in tap water for LMn$_x$O$_y$RGO, LMn$_x$O$_y$NH$_2$RGO, HMn$_x$O$_y$RGO and HMn$_x$O$_y$NH$_2$RGO anodes.

| Sponges | Current density (A m$^{-2}$) | Mn$^{2+}$ (mg L$^{-1}$) |
|---|---|---|
| LMn$_x$O$_y$RGO | 20 | 0.08 ± 0.01 |
| | 29 | 0.20 ± 0.02 |
| LMn$_x$O$_y$NH$_2$RGO | 20 | 0.02 ± 0.01 |
| | 29 | 0.21 ± 0 |
| HMn$_x$O$_y$RGO | 20 | 0.47 ± 0.01 |
| | 29 | 0.25 ± 0.01 |
| HMn$_x$O$_y$NH$_2$RGO | 20 | 0.12 ± 0.01 |
| | 29 | 0.01 ± 0 |

The release of Mn was measured to assess the stability of the graphene sponge anodes during the application of current (**Table 1**). The anodes with the low concentration of Mn$_x$O$_y$ and Mn$_x$O$_y$NH$_2$ both exhibited the same tendency of promoted Mn$^{2+}$ release with the increase in current density. For example, for LMn$_x$O$_y$RGO anode, the release of Mn$^{2+}$ was increased from 0.08 mg L$^{-1}$ at 20 A m$^{-2}$, to 0.20 mg L$^{-1}$ of Mn$^{2+}$ at 29 A m$^{-2}$. On the other hand, anodes with a high concentration of Mn$_x$O$_y$ and Mn$_x$O$_y$NH$_2$ show the opposite trend, with the lower Mn$^{2+}$ release observed at higher current density. The release is caused by MnIII instability at pH <9 which disproportionates into Mn$^{2+}$ and MnIV (2MnIII → Mn$^{2+}$ + MnO$_2$) [44]. However, when higher current densities are applied (e.g., 29 A m$^{-2}$) Mn$^{2+}$ is oxidized into the coating, thus enabling the regeneration of manganese oxide in the sponge [46]. The stability of the material is further enhanced with amino doping, as the formation of N-Mn bonds stabilizes the MnIII [44]. Therefore, the highest electrochemical stability was observed for the $HMn_xO_yNH_2RGO$ anode at 29 A m$^{-2}$, where barely any manganese was detected in the effluent (i.e., 0.01 mg L$^{-1}$). The WHO guideline concentration for manganese in drinking water is 0.08 mg L$^{-1}$ [47]. Additionally, there was no formation of permanganate in the effluent, which was proven by UV-Vis spectroscopy (**Figure S12**). These results demonstrate that the employed amino doping of the $Mn_xO_y$ greatly contributed to the stability of the manganese oxide incorporated into the RGO coating. Furthermore, UV-Vis spectroscopy measurements of GO and RGO in the reactor effluent at 14, 20 and 29 A m$^{-2}$ of applied current density showed that they were below the detection limit (0.0176 mg L$^{-1}$) (**Figure S12**). This may be due to the formation of covalent bonds (i.e., Si-O-C) between the RGO and $SiO_2$ of the mineral wool [48]. The analysis of the total organic carbon (TOC) showed 3.7 and 3.2 mg L$^{-1}$ of TOC in the influent and effluent treated at 29 A m$^{-2}$, respectively.

To compare the disinfection performance of the most electrochemically stable $HMn_xO_yNH_2RGO$ anode in tap water and synthetic electrolyte of the same conductivity, experiments were also performed in 5 mM phosphate buffer (0.47 mS cm$^{-1}$). The removal of *E. coli* at 29 A m$^{-2}$ was 1.4 log in 5 mM phosphate buffer and 2.0 log in tap water (**Figure S13**). The difference in removal is associated with the blockage of active sites by the large and strongly hydrated phosphate ions, as they are not able to penetrate the nanosheets of RGO [49]. In addition, Mn phosphate precipitate was observed after the experiments, hence the amino doping is not preventing the release of manganese in the presence of high concentrations of phosphate.

The potential regrowth of *E. coli* was assessed by incubating the electrochemically treated samples for 18 h at 25 and 37 °C (**Figure 5**). These temperatures were chosen because 25 °C is a common room temperature, whereas 37 °C is the optimum growth temperature for *E. coli* and it is often employed in the reactivation experiments [50]. The concentration of *E. coli* after the overnight incubation decreased in all experiments, thus suggesting that the applied treatment caused an irreversible damage to *E. coli* cells. For example, HMn$_x$O$_y$NH$_2$RGO inactivated 2 log of *E. coli* during the electrochemical treatment at 29 A m$^{-2}$, which increased to 2.7 and 4.6 log when storing the samples at 25 and 37°C, respectively. The enhanced inactivation of *E. coli* during storage at a higher temperature can be explained by the lower stability of the cell membrane, which becomes more susceptible to damage and leaking of the intra-cellular material [51]. The overall removal, as a sum of the removal caused by the electrochemical treatment and the storage effect, was enhanced for higher applied current densities, due to the more pronounced initial inactivation and damage to the cell walls caused by the electroporation. Thus, the highest overall *E. coli* removal of 4.7 log and 4.6 log were achieved by the HMn$_x$O$_y$RGO and the HMn$_x$O$_y$NH$_2$RGO, respectively, at 29 A m$^{-2}$ and storage at 37°C. LMn$_x$O$_y$RGO inactivated 3 log and LMn$_x$O$_y$NH$_2$RGO 3.6 log, in the same experimental conditions. Thus, even though the one-pass electrochemical treatment achieved up to 2.2-2.7 log removal of *E. coli*, this was further increased to 4.6-4.7 log removal due to the continued cell dye off during storage. It should be noted here that no attempt was made in this study to optimize the storage time necessary for more complete inactivation of *E. coli* cells, and shorter storage times may be sufficient to reach a similar disinfection performance.

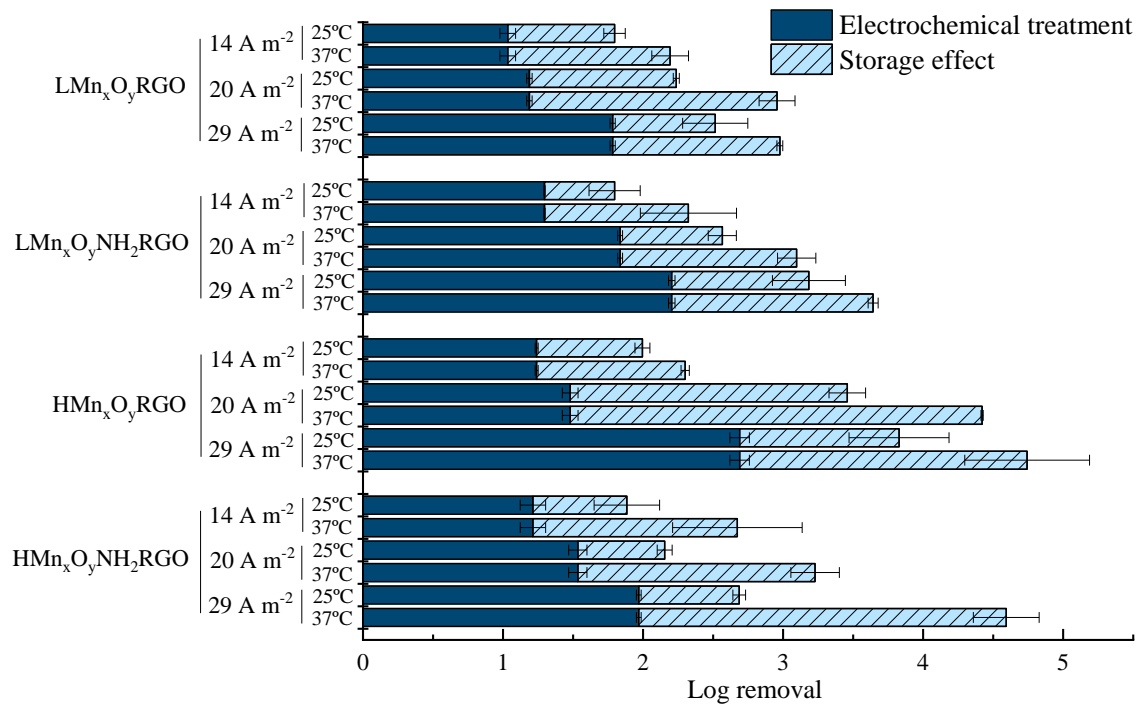

**Figure 5.** Inactivation of *E. coli* ATCC 25922 after the electrochemical treatment of tap water at three applied current densities (14, 20 and 29 A m$^{-2}$) using LMn$_x$O$_y$RGO, LMn$_x$O$_y$NH$_2$RGO, HMn$_x$O$_y$RGO and HMn$_x$O$_y$NH$_2$RGO anode versus the NRGO cathode, and after 18 h storage at 25°C and 37°C.

### 3.4. Electrochemical disinfection with intermittent current

Disinfection experiments were performed using the HMn$_x$O$_y$NH$_2$RGO anode, selected due to its high electrochemical stability, in an intermittent current application mode anode, with asymmetrical 75s ON- 30s OFF pulses at 29 A m$^{-2}$. The duration of the ON and OFF cycles was selected based on our previous study [16]. Under these conditions, 1.53 log removal of *E. coli* was achieved, somewhat lower compared with the continuous current application at the same current density (1.97 log) (**Figure 6**). Although the system operated with the intermittent current did not achieve the same inactivation of *E. coli* as with the continuous current, it still displayed some effect of the pseudo-capacitance of the graphene sponge electrode as 29% reduction in the ON time led to 23% reduction in the disinfection performance (*i.e.*, from 1.97 to 1.53 log). Furthermore, electrochemically treated samples were stored, and the inactivation of *E. coli* increased to 1.9 and 2.8 log at 25 and 37ºC, respectively, for an intermittent current mode (**Figure S14**), indicating that the damage caused by the electroporation was sufficient to not only prevent the cell regrowth but also to cause a further loss in viability. Nevertheless, the inactivation of *E. coli* after storage was still lower after the application of the intermittent current in comparison to the continuous current, with an overall *E. coli* removal of 2.8 and 4.6 log, respectively, for the samples stored at 37 ºC (**Figure 5**). The switching from continuous to intermittent current lowered the energy consumption from 1.25 to 0.89 kWh m$^{-3}$, respectively. The optimization of the duration of ON and OFF cycles was out of scope of the present study; yet it is likely that the application of intermittent current in terms of cycle duration, symmetry, pulse shape (*e.g.*, sharp *vs* gradual step change) could be further fine-tuned to achieve higher inactivation of *E. coli*.

### 3.5. Mechanism of *E. coli* inactivation

**Figure 7** shows a SEM image of *E. coli* cells before (**Fig. 7A**) and after (**Fig. 7B-D**) the electrochemical treatment. As evidenced by the images, treated cells showed protrusions in the cell membrane that resemble intracellular material leaking through numerous pores. Altogether, images suggest that the loss of cell viability was probably caused by electroporation, which is the application of an electric field to a cell to increase the permeability of the membrane [52]. The negatively charged *E. coli* cells were electrosorbed to the positively charged graphene sponge anode and the applied electric field caused damage to the cell membrane that resulted in a viability loss. As shown in the reactivation experiments, the damage caused to cells was irreversible, preventing their reactivation and further regrowth. Ozone and hydroxyl radicals are reported of being capable to inactivate *E. coli*, whereas hydrogen peroxide is usually inefficient [53]. Our previous study showed that the addition of methanol as a radical scavenger to determine the contribution of ·OH to the removal of *E. coli* at graphene sponge electrodes led to a higher inactivation of *E. coli* cells, contrary to the expected effect of methanol (i.e., worsened disinfection due to the presence of methanol) [16]. This was due to the enhanced toxicity of methanol in the presence of electric field, and its enhanced penetration inside the electroporated *E. coli* cells. The concentration of the produced oxidants was similar in the four studied systems (**Figure S11**), thus the differences in removal between systems were associated with different degrees of electroporation.

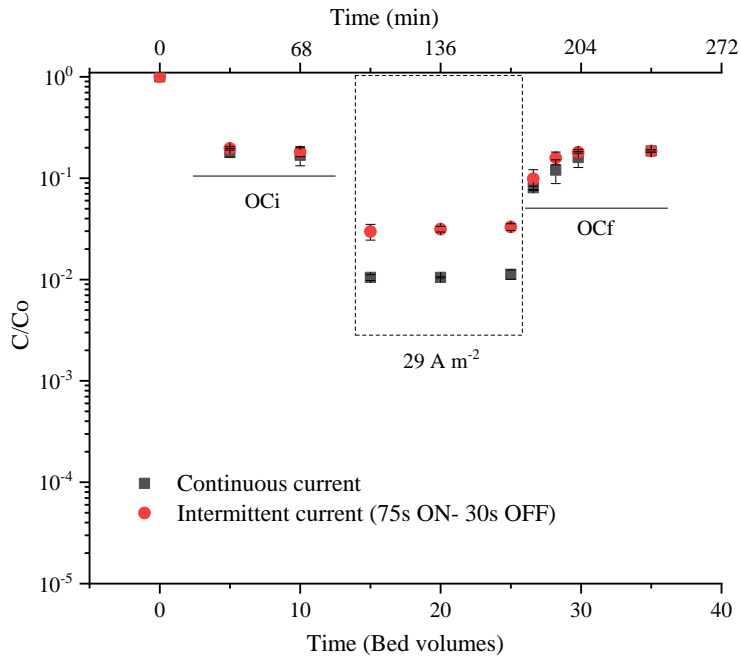

**Figure 6.** Electrochemical inactivation of *E. coli* ATCC 25922 in tap water using $HMn_xO_yNH_2RGO$ anode at 29 A m$^{-2}$ in continuous and intermittent current application mode, with 75s ON - 30s OFF cycles.

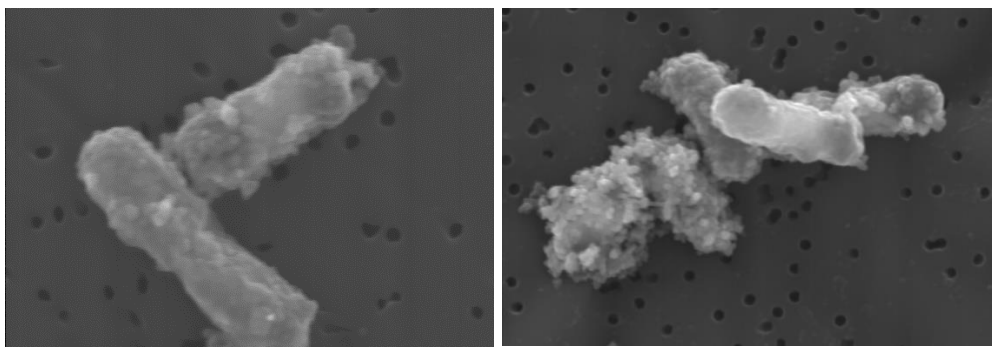

**Figure 7.** SEM images of *E. coli* ATCC 25922 cells **(A)** before and **(B-D)** after the electrochemical treatment.

Electroporation typically requires strong electric fields of ~$10^5$ V cm$^{-1}$ [54], whereas the potential gradient in our system was 3.6 - 4.4 V cm$^{-1}$. Low-voltage electroporation at graphene sponge anode was also reported for a different *E. coli* strain (ATCC 700078) in our previous study [16], and was assigned to the nanostructure of the RGO coating that enhances the electric field at the anode surface. Similar phenomena was observed previously at other nanostructured anodes based on carbon nanotubes (CNTs) [55] and copper oxide nanowires [56]. Using the same *E. coli* strain that we used in this work (ATCC 25922), Hong et al. report a silver nanowire carbon fiber cloth that inactivates around 3.3 log of *E. coli* when applying 6V of anodic potential with a flow rate of 5 mL min$^{-1}$ in 10 mM NaCl [57]. In our previous study, N-doped graphene sponge anode achieved 5.5 log removal of *E. coli* ATCC 700078 in tap water at 58 A m$^{-2}$ [16]. In the present study, however, higher electrical resistance of the synthesized manganese oxide-functionalized graphene sponge anode did not allow the application of current densities higher than 29 A m$^{-2}$ in tap water. Furthermore, strain ATCC 700078 has an altered external membrane structure [58], which may facilitate its damage by electroporation compared to the strain ATCC 25922 employed in the present work. Nevertheless, given the low cost of the developed graphene sponge electrodes and relatively low energy consumption when treating low conductivity tap water (1.2-1.6 kWh m$^{-3}$) (**Table S7**), more complete disinfection of tap water could likely be achieved by using multiple treatment units, as well as reactors with optimized hydrodynamics. The applied treatment caused irreversible damage to *E. coli* cells that prevented their further reactivation. Although the present study investigated only the impact of overnight storage (*i.e.*, 18 h), shorter storage times may also be sufficient to achieve the same inactivation rates.

4. Conclusions

Manganese oxide-functionalized graphene sponge electrodes were demonstrated to inactivate the resistant *E. coli* bacteria in real tap water, without the formation of toxic chlorinated by-products. The interaction between the RGO and Mn decreased the oxidation state of Mn (Mn II and Mn III) and favored the formation of oxygen vacancies, which enhanced the electrocatalytic and capacitive properties of the material. Comparison of the $Mn_xO_y$-functionalized graphene sponge anodes with the non-functionalized RGO anode demonstrated that higher $Mn_xO_y$ loading led to an enhanced electrochemical disinfection performance. In real tap water, $HMn_xO_yRGO$ anode achieved the highest removal of *E. coli* (2.7 log removal at 29 A m$^{-2}$) due to higher anode potential and potential gradient, however $Mn^{2+}$ release was observed during the treatment, with up to 0.25 mg L$^{-1}$ of total Mn detected in the treated effluent. $HMn_xO_yNH_2RGO$ anode inactivated 2 log of *E. coli* but with the minimal release of $Mn^{2+}$ (0.01 mg L$^{-1}$), demonstrating the stabilizing role of the amino doping of manganese oxide.

After storing the electrochemically treated samples for 18 h at 25 and 37 °C, 2.7 log and 4.6 log of *E. coli* removal were achieved, respectively, indicating that the treatment caused irreparable damage to the cell membranes *via* electrosorption of the bacteria at the anode and disruption of the membrane by low-voltage electroporation. Furthermore, we report for the first time the crucial impact of the storage temperature on further bacterial inactivation when their damage is caused by electroporation, as 37ºC typically applied in the regrowth studies as the optimum temperature for *E. coli* growth led to a higher bacterial cells dye off compared with the lower temperatures of 25ºC.

Intermittent current of 75s ON – 30s OFF was applied to take advantage of the pseudo-capacitance of the sponge. 1.53 log of *E. coli* were inactivated during the treatment, increasing to 2.8 log after storing the samples 18h at 37ºC. Although the pseudo-capacitance of the graphene sponge anode had a limited contribution to the electrochemical disinfection of tap water, the duration, symmetry, and shape of the current cycles could be further optimized to reach higher *E. coli* inactivation and with lower energy consumption. In summary, this study shows the potential of the low-cost $Mn_xO_y$-$NH_2$-functionalized graphene sponge anodes for chlorine-free disinfection of real tap water.

**Acknowledgements**

The authors would like to acknowledge ERC Starting Grant project ELECTRON4WATER (Three-dimensional nanoelectrochemical systems based on low-cost reduced graphene oxide: the next generation of water treatment systems) [project number 714177]. ICRA researchers thank funding from CERCA program and the support from the Economy and Knowledge Department of the Catalan Government through the Consolidated Research Group program to [ICRA-TECH – 2021 SGR 0128] and [2021 SGR 01282 ICRA-ENV].

**References**

[1] World Health Organisation; UN-Water, UN-Water Global Analysis and Assessment of Sanitation and Drinking-Water (GLAAS), 2017. https://doi.org/10.1016/0002-9149(58)90231-5.

[2] J. Sanchís, A. Jaén-Gil, P. Gago-Ferrero, E. Munthali, M.J. Farré, Characterization of organic matter by HRMS in surface waters: Effects of chlorination on molecular fingerprints and correlation with DBP formation potential, Water Res. 176 (2020). https://doi.org/10.1016/j.watres.2020.115743.

[3] M.T. Guo, C. Kong, Antibiotic resistant bacteria survived from UV disinfection: Safety concerns on genes dissemination, Chemosphere. 224 (2019) 827–832. https://doi.org/10.1016/j.chemosphere.2019.03.004.

[4] J. Wang, H. Chen, Catalytic ozonation for water and wastewater treatment: Recent advances and perspective, Sci. Total Environ. 704 (2020) 135249. https://doi.org/10.1016/j.scitotenv.2019.135249.

[5] S. Hand, R.D. Cusick, Electrochemical Disinfection in Water and Wastewater Treatment: Identifying Impacts of Water Quality and Operating Conditions on Performance, Environ. Sci. Technol. 55 (2021) 3470–3482. https://doi.org/10.1021/acs.est.0c06254.

[6] B.P. Chaplin, Advantages, disadvantages, and future challenges of the use of electrochemical technologies for water and wastewater treatment, Elsevier Inc., 2018. https://doi.org/10.1016/B978-0-12-813160-2.00017-1.

[7] V. Schmalz, T. Dittmar, D. Haaken, E. Worch, Electrochemical disinfection of biologically treated wastewater from small treatment systems by using boron-doped diamond (BDD) electrodes - Contribution for direct reuse of domestic wastewater, Water Res. 43 (2009) 5260–5266. https://doi.org/10.1016/j.watres.2009.08.036.

[8] E. Mordačíková, M. Vojs, K. Grabicová, M. Marton, P. Michniak, V. Řeháček, A. Bořík, R. Grabic, J. Bruncko, T. Mackuľak, A. Vojs Staňová, Influence of boron doped diamond electrodes properties on the elimination of selected pharmaceuticals from wastewater, J. Electroanal. Chem. 862 (2020). https://doi.org/10.1016/j.jelechem.2020.114007.

[9] L. Guo, K. Ding, K. Rockne, M. Duran, B.P. Chaplin, Bacteria inactivation at a sub-stoichiometric titanium dioxide reactive electrochemical membrane, J. Hazard. Mater. 319 (2016) 137–146. https://doi.org/10.1016/j.jhazmat.2016.05.051.

[10] N. Ghasemian, Saloumeh; Asadishad, Bahareh; Omanovic, Sasha; Tufenkji, Electrochemical disinfection of bacteria-laden water using antimony-doped tin-tungsten-oxide electrodes, Water Res. 126 (2017) 299–307.

[11] X. Cong, J. Bao, Preparation of a PbO2 electrode with graphene interlayer and for electrochemical oxidation of doxycycline, Int. J. Electrochem. Sci. 15 (2020) 4352–4367. https://doi.org/10.20964/2020.05.16.

[12] J. Radjenovic, D.L. Sedlak, Challenges and Opportunities for Electrochemical Processes as Next-Generation Technologies for the Treatment of Contaminated Water, Environ. Sci. Technol. 49 (2015) 11292–11302. https://doi.org/10.1021/acs.est.5b02414.

[13] M.E.H. Bergmann, J. Rollin, T. Iourtchouk, The occurrence of perchlorate during drinking water electrolysis using BDD anodes, Electrochim. Acta. 54 (2009) 2102–2107. https://doi.org/10.1016/j.electacta.2008.09.040.

[14] M.H. Lin, D.M. Bulman, C.K. Remucal, B.P. Chaplin, Chlorinated Byproduct Formation during the Electrochemical Advanced Oxidation Process at Magnéli Phase Ti4O7 Electrodes, Environ. Sci. Technol. 54 (2020) 12673–12683.

https://doi.org/10.1021/acs.est.0c03916.

[15]  L. Baptista-Pires, G.F. Norra, J. Radjenovic, Graphene-based sponges for electrochemical degradation of persistent organic contaminants, Water Res. 203 (2021). https://doi.org/10.1016/j.watres.2021.117492.

[16]  G.-F. Norra, L. Baptista-Pires, E.C. Lumbaque, C.M. Borrego, J. Radjenovic, Chlorine-free electrochemical disinfection using graphene sponge electrodes, Chem. Eng. J. (2021) 132772.

[17]  C. Guan, C. Guan, Q. Guo, R. Huang, J. Duan, Z. Wang, X. Wei, J. Jiang, Enhanced oxidation of organic contaminants by Mn(VII) in water, Water Res. 226 (2022) 119265. https://doi.org/10.1016/j.watres.2022.119265.

[18]  T.S. Sreeprasad, S.M. Maliyekkal, K.P. Lisha, T. Pradeep, Reduced graphene oxide-metal/metal oxide composites: Facile synthesis and application in water purification, J. Hazard. Mater. 186 (2011) 921–931. https://doi.org/10.1016/j.jhazmat.2010.11.100.

[19]  H. Wu, X. Xu, L. Shi, Y. Yin, L.C. Zhang, Z. Wu, X. Duan, S. Wang, H. Sun, Manganese oxide integrated catalytic ceramic membrane for degradation of organic pollutants using sulfate radicals, Water Res. 167 (2019). https://doi.org/10.1016/j.watres.2019.115110.

[20]  A. Massa, S. Hernández, A. Lamberti, C. Galletti, N. Russo, D. Fino, Electro-oxidation of phenol over electrodeposited $MnO_x$ nanostructures and the role of a $TiO_2$ nanotubes interlayer, Appl. Catal. B Environ. 203 (2017) 270–281. https://doi.org/10.1016/j.apcatb.2016.10.025.

[21]  Y. Hu, Y. Wu, J. Wang, Manganese-Oxide-Based Electrode Materials for Energy Storage Applications: How Close Are We to the Theoretical Capacitance?, Adv. Mater. 30 (2018) 1–21. https://doi.org/10.1002/adma.201802569.

[22]  J. Xu, Y. Li, M. Qian, J. Pan, J. Ding, B. Guan, Amino-functionalized synthesis of $MnO_2$-$NH_2$-GO for catalytic ozonation of cephalexin, Appl. Catal. B Environ. 256 (2019). https://doi.org/10.1016/j.apcatb.2019.117797.

[23]  Y. Jin, Y. Shi, R. Chen, X. Chen, X. Zheng, Y. Liu, Electrochemical disinfection using a modified reticulated vitreous carbon cathode for drinking water treatment, Chemosphere. 215 (2019) 380–387. https://doi.org/10.1016/j.chemosphere.2018.10.057.

[24]  Y. Li, M. Yang, X. Zhang, J. Jiang, J. Liu, C.F. Yau, N.J.D. Graham, X. Li, Two-step chlorination: A new approach to disinfection of a primary sewage effluent, Water Res. 108 (2017) 339–347. https://doi.org/10.1016/j.watres.2016.11.019.

[25]  R.A. Baga, Ayad N.; Johnson, G.R. Alastair; Nazhat, Najdat B.; Saadalla-Nazhat, A simple spectrophotometric determination of hydrogen peroxide at low concetrations in aqueous solution, Anal. Chim. Acta. 204 (1988) 349–353.

[26]  S. Nayak, B.P. Chaplin, Fabrication and characterization of porous, conductive, monolithic $Ti_4O_7$ electrodes, Electrochim. Acta. 263 (2018) 299–310. https://doi.org/10.1016/j.electacta.2018.01.034.

[27]  X. Duan, J. Yang, H. Gao, J. Ma, L. Jiao, W. Zheng, Controllable hydrothermal synthesis of manganese dioxide nanostructures: Shape evolution, growth mechanism and electrochemical properties, CrystEngComm. 14 (2012) 4196–4204. https://doi.org/10.1039/c2ce06587h.

[28]  G. Park, L. Bartolome, K.G. Lee, S.J. Lee, D.H. Kim, T.J. Park, One-step sonochemical synthesis of a graphene oxide-manganese oxide nanocomposite for catalytic glycolysis of poly(ethylene terephthalate), Nanoscale. 4 (2012) 3879–3885. https://doi.org/10.1039/c2nr30168g.

[29] A.T. Chidembo, S.H. Aboutalebi, K. Konstantinov, C.J. Jafta, H.K. Liu, K.I. Ozoemena, In situ engineering of urchin-like reduced graphene oxide–Mn2O3–Mn3O4 nanostructures for supercapacitors, RSC Adv. 4 (2014) 886–892. https://doi.org/10.1039/C3RA44973D.

[30] P. Hosseini-Benhangi, C.H. Kung, A. Alfantazi, E.L. Gyenge, Controlling the Interfacial Environment in the Electrosynthesis of MnOx Nanostructures for High-Performance Oxygen Reduction/Evolution Electrocatalysis, ACS Appl. Mater. Interfaces. 9 (2017) 26771–26785. https://doi.org/10.1021/acsami.7b05501.

[31] L. Zhong Zhao, V. Young, XPS studies of carbon supported films formed by the resistive deposition of manganese, J. Electron Spectros. Relat. Phenomena. 34 (1984) 45–54. https://doi.org/10.1016/0368-2048(84)80058-4.

[32] C. Hou, Y. Hou, Y. Fan, Y. Zhai, Y. Wang, Z. Sun, R. Fan, F. Dang, J. Wang, Oxygen vacancy derived local build-in electric field in mesoporous hollow Co3O4 microspheres promotes high-performance Li-ion batteries, J. Mater. Chem. A. 6 (2018) 6967–6976. https://doi.org/10.1039/c8ta00975a.

[33] M. Chigane, M. Ishikawa, Manganese Oxide Thin Film Preparation by Potentiostatic Electrolyses and Electrochromism, J. Electrochem. Soc. 147 (2000) 2246. https://doi.org/10.1149/1.1393515.

[34] N. Liu, X. Wu, Y. Yin, A. Chen, C. Zhao, Z. Guo, L. Fan, L. Fan, N. Zhang, N. Zhang, Constructing the Efficient Ion Diffusion Pathway by Introducing Oxygen Defects in Mn2O3for High-Performance Aqueous Zinc-Ion Batteries, ACS Appl. Mater. Interfaces. 12 (2020) 28199–28205. https://doi.org/10.1021/acsami.0c05968.

[35] A. Hodgson, S. Haq, Water adsorption and the wetting of metal surfaces, Surf. Sci. Rep. 64 (2009) 381–451. https://doi.org/10.1016/j.surfrep.2009.07.001.

[36] J. Qu, L. Shi, C. He, F. Gao, B. Li, Q. Zhou, H. Hu, G. Shao, X. Wang, J. Qiu, Highly efficient synthesis of graphene/MnO2 hybrids and their application for ultrafast oxidative decomposition of methylene blue, Carbon N. Y. 66 (2014) 485–492. https://doi.org/10.1016/j.carbon.2013.09.025.

[37] A. Sumboja, C.Y. Foo, X. Wang, P.S. Lee, Large areal mass, flexible and free-standing reduced graphene oxide/manganese dioxide paper for asymmetric supercapacitor device, Adv. Mater. 25 (2013) 2809–2815. https://doi.org/10.1002/adma.201205064.

[38] T. He, X. Zeng, S. Rong, The controllable synthesis of substitutional and interstitial nitrogen-doped manganese dioxide: The effects of doping sites on enhancing the catalytic activity, J. Mater. Chem. A. 8 (2020) 8383–8396. https://doi.org/10.1039/d0ta01346c.

[39] A. Chakravarty, D. Sengupta, B. Basu, A. Mukherjee, G. De, MnO2 nanowires anchored on amine functionalized graphite nanosheets: Highly active and reusable catalyst for organic oxidation reactions, RSC Adv. 5 (2015) 92585–92595. https://doi.org/10.1039/c5ra17777d.

[40] F. Gao, X. Tang, H. Yi, C. Chu, N. Li, J. Li, S. Zhao, In-situ DRIFTS for the mechanistic studies of NO oxidation over A-MnO2, B-MnO2 and Γ-MnO2 catalysts, Chem. Eng. J. 322 (2017) 525–537. https://doi.org/10.1016/j.cej.2017.04.006.

[41] L. Tian, X. Zhai, X. Wang, J. Li, Z. Li, Advances in manganese-based oxides for oxygen evolution reaction, J. Mater. Chem. A. 8 (2020) 14400–14414. https://doi.org/10.1039/d0ta05116k.

[42] F. Perreault, A.F. De Faria, S. Nejati, M. Elimelech, Antimicrobial Properties of Graphene Oxide Nanosheets: Why Size Matters, ACS Nano. 9 (2015) 7226–7236. https://doi.org/10.1021/acsnano.5b02067.

[43] S. Liu, T.H. Zeng, M. Hofmann, E. Burcombe, J. Wei, R. Jiang, J. Kong, Y. Chen, Antibacterial activity of graphite, graphite oxide, graphene oxide, and reduced graphene oxide: Membrane and oxidative stress, ACS Nano. 5 (2011) 6971–6980. https://doi.org/10.1021/nn202451x.

[44] T. Takashima, K. Hashimoto, R. Nakamura, Inhibition of charge disproportionation of MnO 2 electrocatalysts for efficient water oxidation under neutral conditions, J. Am. Chem. Soc. 134 (2012) 18153–18156. https://doi.org/10.1021/ja306499n.

[45] S. Richardson, A. Iles, J.M. Rotchell, T. Charlson, A. Hanson, M. Lorch, N. Pamme, Citizen-led sampling to monitor phosphate levels in freshwater environments using a simple paper microfluidic device, PLoS One. 16 (2021) 1–13. https://doi.org/10.1371/journal.pone.0260102.

[46] N. Sergienko, J. Radjenovic, Manganese oxide coated TiO2 nanotube-based electrode for efficient and selective electrocatalytic sulfide oxidation to colloidal sulfur, Appl. Catal. B Environ. 296 (2021) 120383. https://doi.org/10.1016/j.apcatb.2021.120383.

[47] U. States, E. Protection, E.P. Agency, A. Pharmacology, Manganese in Drinking-Water Manganese in Drinking-Water, 158 (2002) 271–279.

[48] C. Hintze, K. Morita, R. Riedel, E. Ionescu, G. Mera, Facile sol-gel synthesis of reduced graphene oxide/silica nanocomposites, J. Eur. Ceram. Soc. 36 (2016) 2923–2930. https://doi.org/10.1016/j.jeurceramsoc.2015.11.033.

[49] N. Duinslaeger, A. Doni, J. Radjenovic, Impact of supporting electrolyte on electrochemical performance of borophene-functionalized graphene sponge anode and degradation of per- and polyfluoroalkyl substances (PFAS), Water Res. 242 (2023) 120232. https://doi.org/10.1016/j.watres.2023.120232.

[50] A.M. Gorito, J.F.J.R. Pesqueira, N.F.F. Moreira, A.R. Ribeiro, M.F.R. Pereira, O.C. Nunes, C.M.R. Almeida, A.M.T. Silva, Ozone-based water treatment (O3, O3/UV, O3/H2O2) for removal of organic micropollutants, bacteria inactivation and regrowth prevention, J. Environ. Chem. Eng. 9 (2021) 10–14. https://doi.org/10.1016/j.jece.2021.105315.

[51] K. Aronsson, U. Rönner, Influence of pH, water activity and temperature on the inactivation of Escherichia coli and Saccharomyces cerevisiae by pulsed electric fields, Innov. Food Sci. Emerg. Technol. 2 (2001) 105–112. https://doi.org/10.1016/S1466-8564(01)00030-3.

[52] Z.Y. Huo, G.Q. Li, T. Yu, C. Feng, Y. Lu, Y.H. Wu, C. Yu, X. Xie, H.Y. Hu, Cell Transport Prompts the Performance of Low-Voltage Electroporation for Cell Inactivation, Sci. Rep. 8 (2018) 1–10. https://doi.org/10.1038/s41598-018-34027-0.

[53] J. Jeong, J.Y. Kim, J. Yoon, The role of reactive oxygen species in the electrochemical inactivation of microorganisms, Environ. Sci. Technol. 40 (2006) 6117–6122. https://doi.org/10.1021/es0604313.

[54] T.Y. Tsong, Electroporation of cell membranes, Biophys. J. 60 (1991) 297–306. https://doi.org/10.1016/S0006-3495(91)82054-9.

[55] Z.Y. Huo, Y. Luo, X. Xie, C. Feng, K. Jiang, J. Wang, H.Y. Hu, Carbon-nanotube sponges enabling highly efficient and reliable cell inactivation by low-voltage electroporation, Environ. Sci. Nano. 4 (2017) 2010–2017. https://doi.org/10.1039/c7en00558j.

[56] Z.Y. Huo, X. Xie, T. Yu, Y. Lu, C. Feng, H.Y. Hu, Nanowire-Modified Three-Dimensional Electrode Enabling Low-Voltage Electroporation for Water Disinfection,
32

Environ. Sci. Technol. 50 (2016) 7641–7649. https://doi.org/10.1021/acs.est.6b01050.

[57] X. Hong, J. Wen, X. Xiong, Y. Hu, Silver nanowire-carbon fiber cloth nanocomposites synthesized by UV curing adhesive for electrochemical point-of-use water disinfection, Chemosphere. 154 (2016) 537–545. https://doi.org/10.1016/j.chemosphere.2016.04.013.

[58] L. Imamovic, M. Misiakou, E. Van Der Helm, G. Panagiotou, M. Otto, A. Sommer, Complete Genome Sequence of Escherichia coli Strain WG5, Genome Announc. 6 (2018). https://doi.org/https://doi.org/10.1128/genomeA.01403-17.

# Supplementary Material

# Manganese oxide-functionalized graphene sponge electrodes for electrochemical chlorine-free disinfection of tap water

*Anna Segues Codina[a,b], Natalia Sergienko[a,b], Carles M. Borrego[a,c], Jelena Radjenovic [a,d*]*

[a] *Catalan Institute for Water Research (ICRA-CERCA), Scientific and Technological Park of the University of Girona, 17003 Girona, Spain*

[b] *University of Girona, Girona, Spain*

[c] *Group of Molecular Microbial Ecology, Institute of Aquatic Ecology, University of Girona, 17011 Girona, Spain*

[d] *Catalan Institution for Research and Advanced Studies (ICREA), Passeig Lluís Companys 23, 08010 Barcelona, Spain*

*\* Corresponding author:*

*Jelena Radjenovic, Catalan Institute for Water Research (ICRA), c/Emili Grahit, 101, 17003 Girona, Spain*

Phone: + 34 972 18 33 80; Fax: +34 972 18 32 48; E-mail: jradjenovic@icra.cat

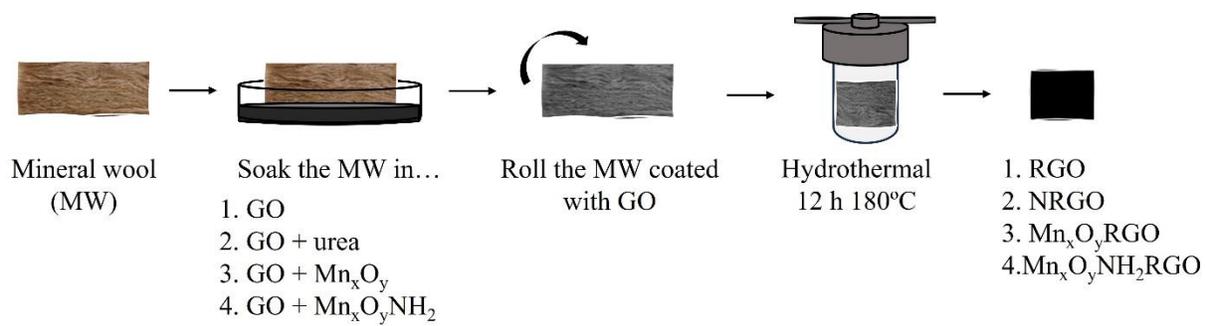

**Figure S1.** Schematic illustration of the graphene-based sponge electrode fabrication.

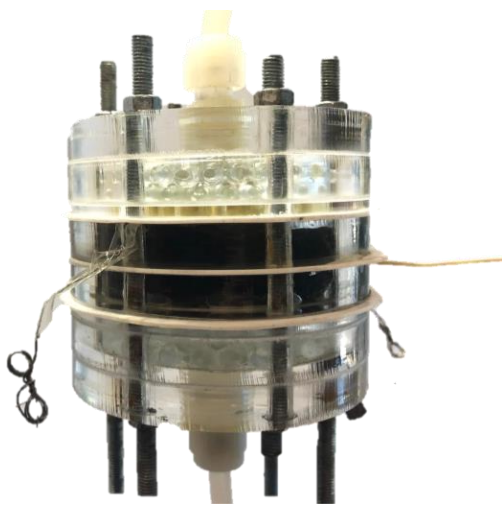

**Figure S2.** One-pass cylindrical flow-through reactor used for the electrochemical disinfection experiments.

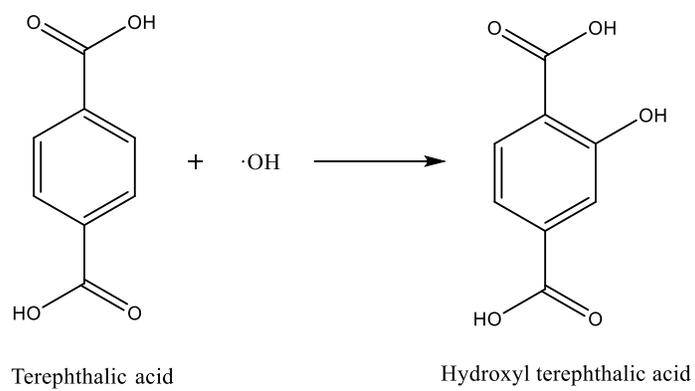

Terephthalic acid → Hydroxyl terephthalic acid

**Figure S3.** Reaction between terephthalic acid (TA) and hydroxyl radicals (·OH), for the determination of ·OH.

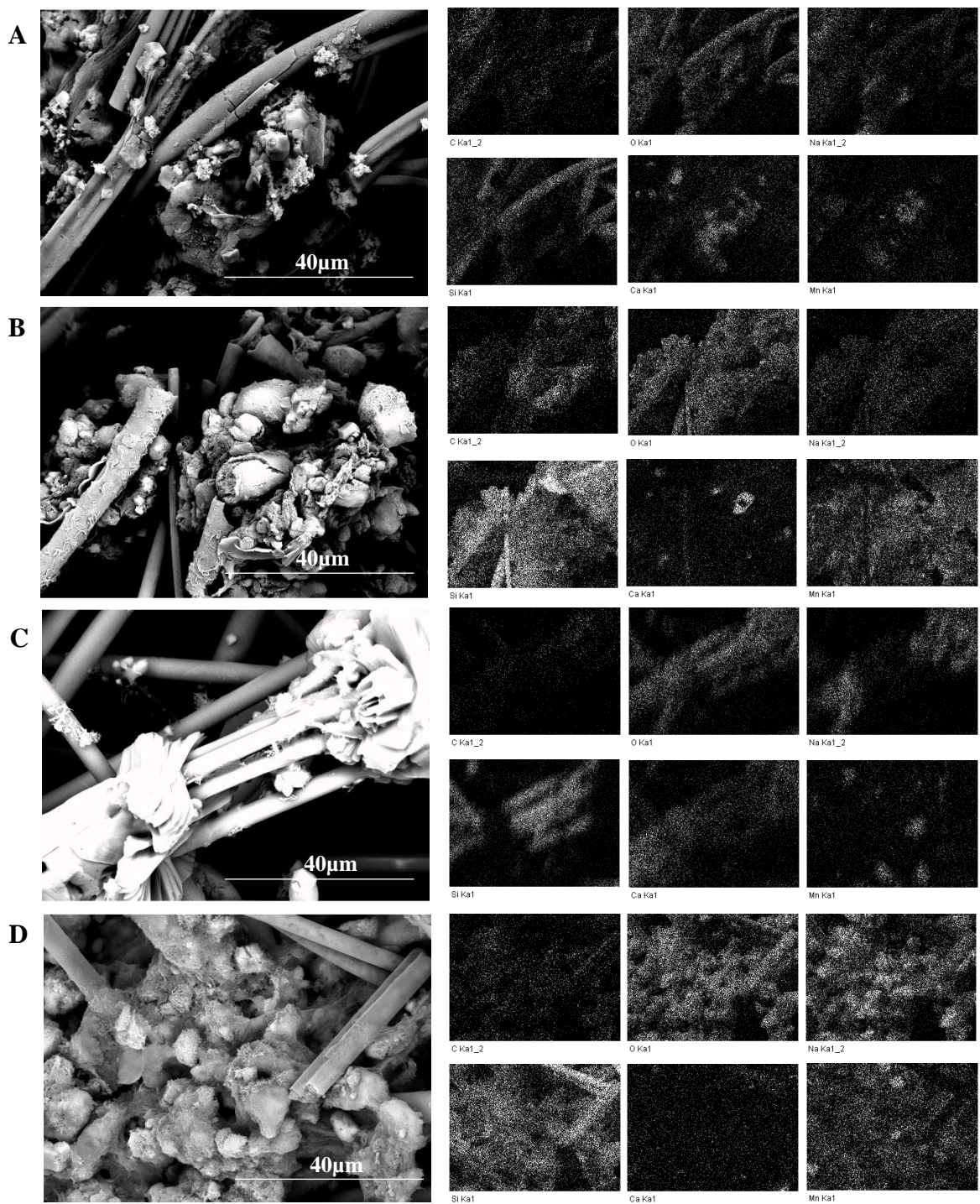

**Figure S4.** Scanning electron microscopy (SEM) images with an element mapping (*i.e.* C, O, Na, Si, Ca and Mn) for **(A)** $LMn_xO_yRGO$, **(B)** $LMn_xO_yNH_2RGO$, **(C)** $HMn_xO_yRGO$, and **(D)** $HMn_xO_yNH_2RGO$.

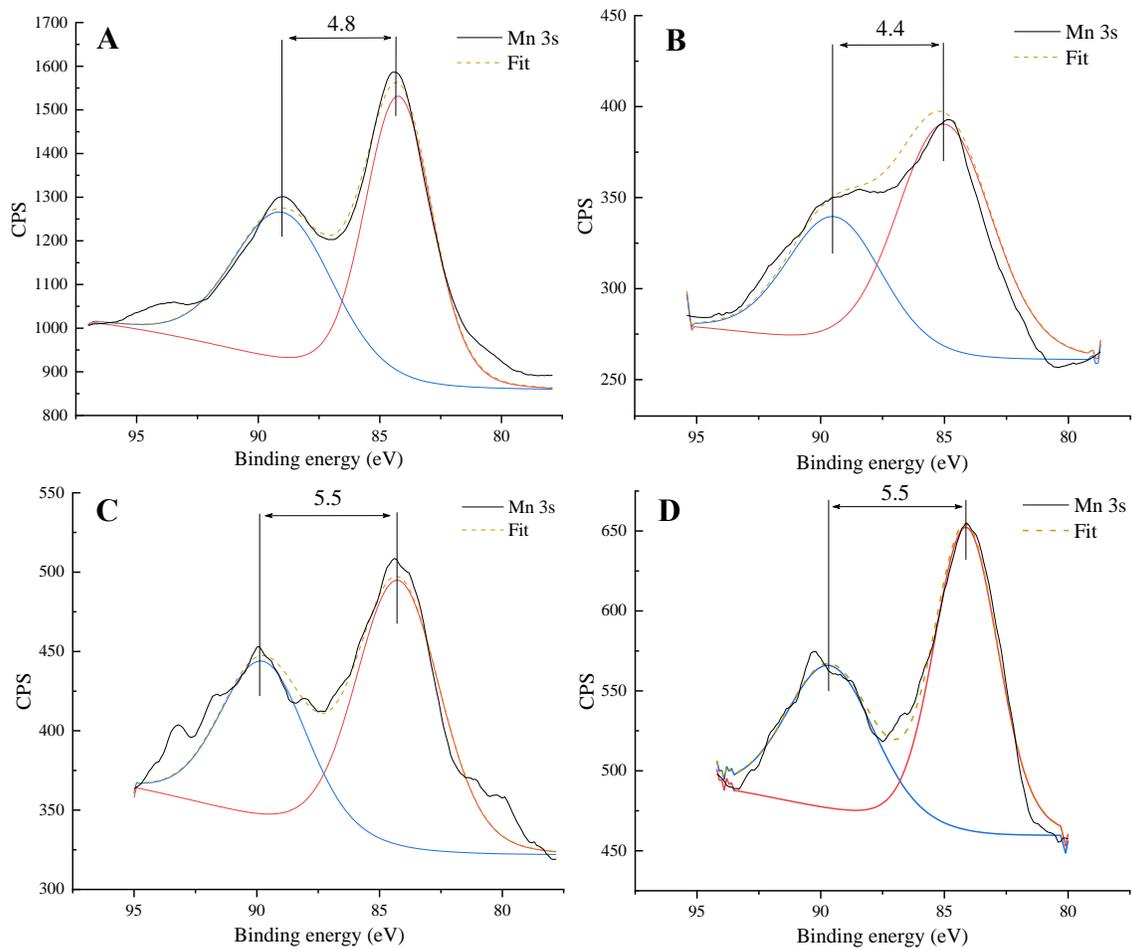

**Figure S5.** Mn 3s spectra of **(A)** $Mn_xO_y$, **(B)** $Mn_xO_yNH_2$, **(C)** $HMn_xO_yRGO$ and **(D)** $HMn_xO_yNH_2RGO$ by X-ray photoelectron spectroscopy (XPS).

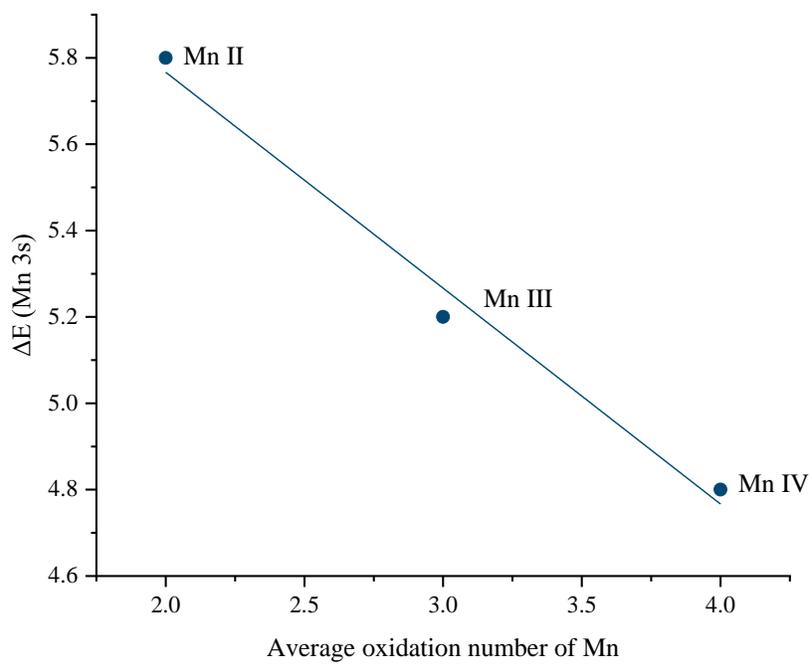

**Figure S6.** Correlation between the separation of Mn 3s peaks (ΔE) and the average oxidation number of Mn, according to the values reported in the literature. [1]

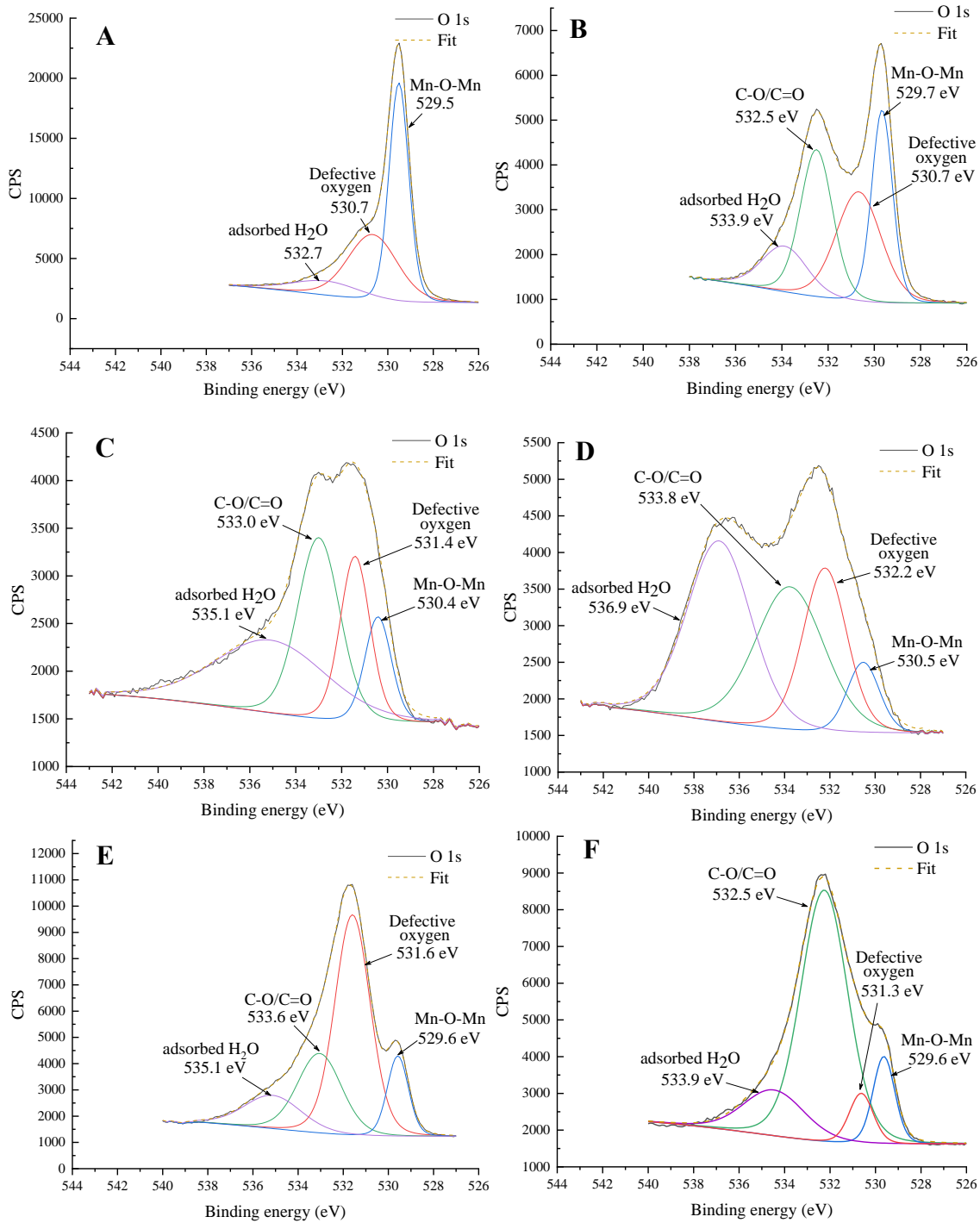

**Figure S7.** O 1s spectra of **(A)** $Mn_xO_y$, **(B)** $Mn_xO_yNH_2$, **(C)** $LMn_xO_yRGO$, **(D)** $LMn_xO_yNH_2RGO$, **(E)** $HMn_xO_yRGO$ and **(F)** $HMn_xO_yNH_2RGO$ obtained by X-ray photoelectron spectroscopy (XPS).

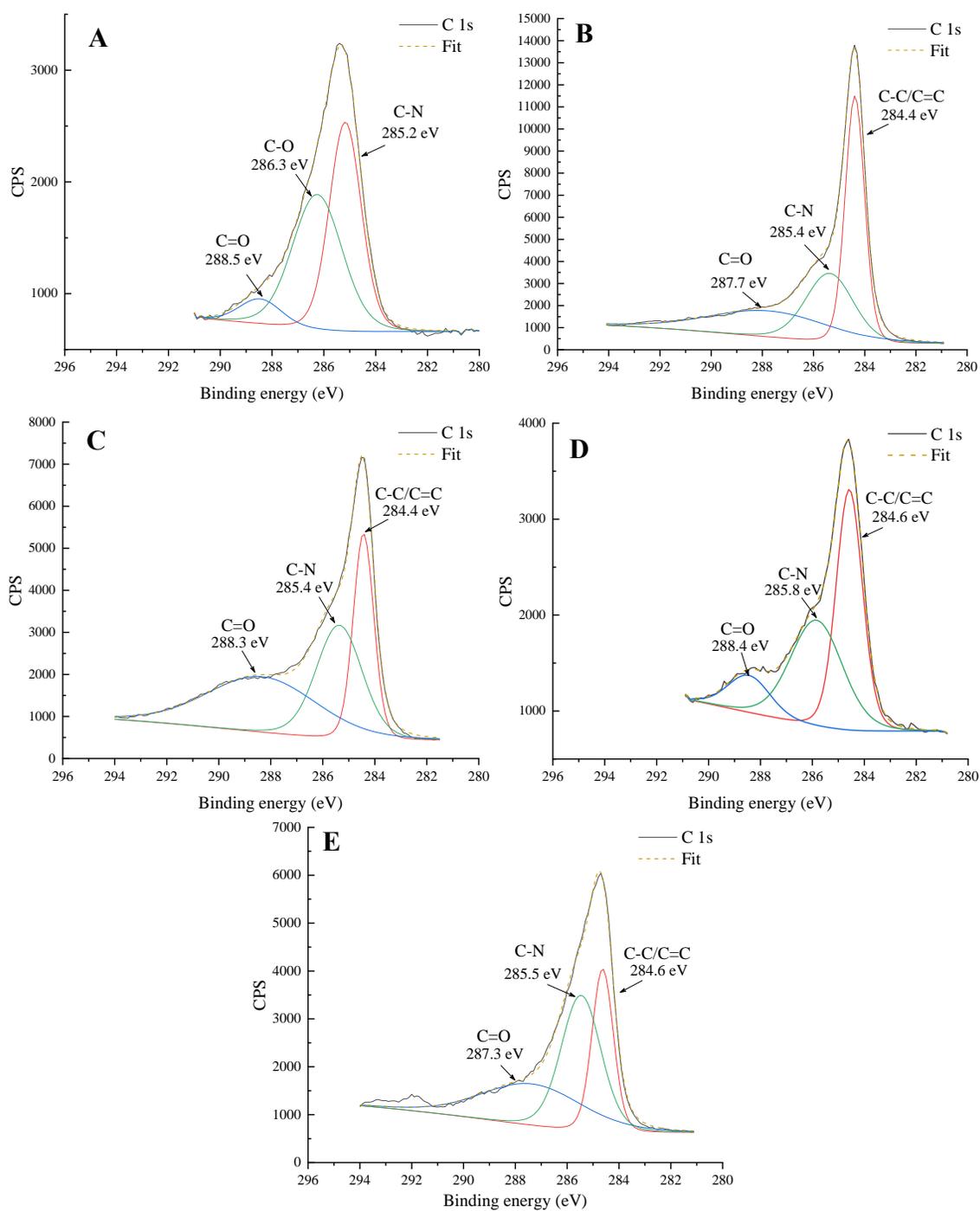

**Figure S8.** C 1s spectra of **(A)** $Mn_xO_yNH_2$, **(B)** $LMn_xO_yRGO$, **(C)** $LMn_xO_yNH_2RGO$, **(D)** $HMn_xO_yRGO$ and **(E)** $HMn_xO_yNH_2RGO$ obtained by X-ray photoelectron spectroscopy (XPS).

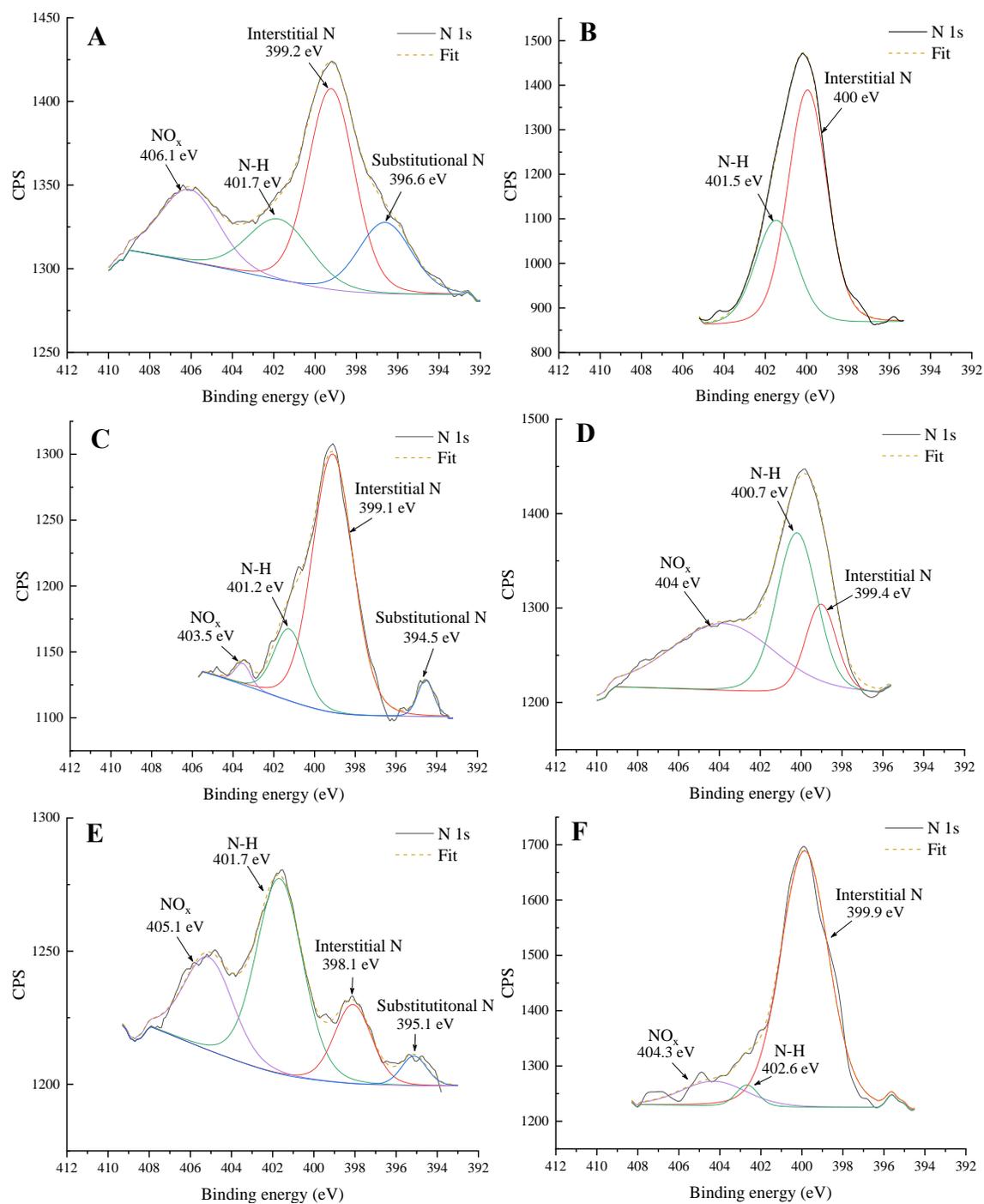

**Figure S9.** N 1s spectra of **(A)** $Mn_xO_y$, **(B)** $Mn_xO_yNH_2$, **(C)** $LMn_xO_yRGO$, **(D)** $LMn_xO_yNH_2RGO$, **(E)** $HMn_xO_yRGO$ and **(F)** $HMn_xO_yNH_2RGO$ obtained by X-ray photoelectron spectroscopy (XPS).

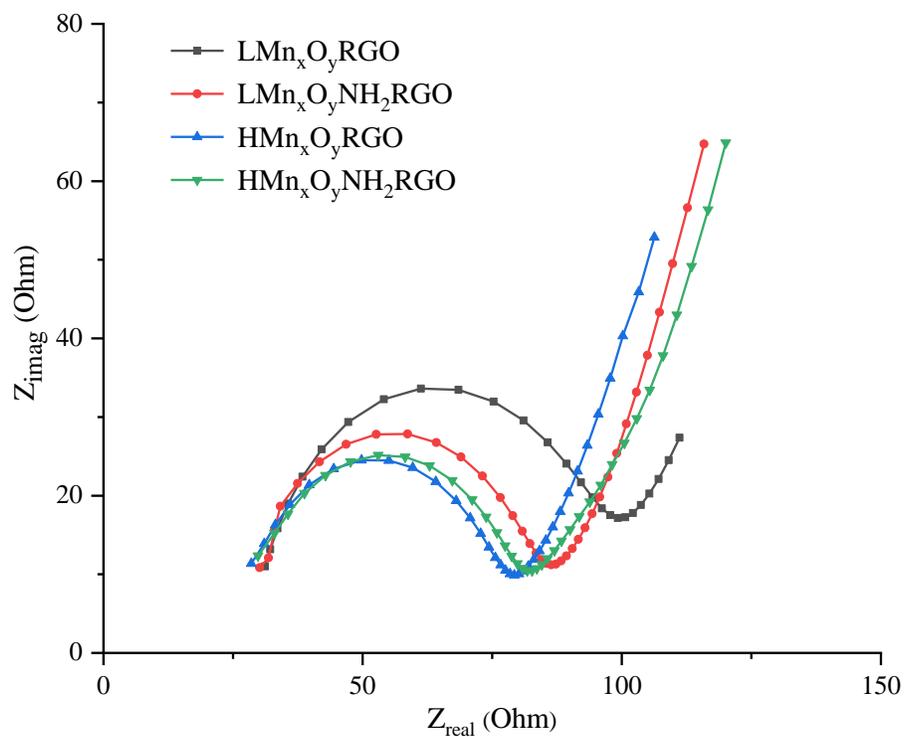

**Figure S10.** Nyquist plots of the $LMn_xO_yRGO$, $LMn_xO_yNH_2RGO$, $HMn_xO_yRGO$ and $HMn_xO_yNH_2RGO$ anode *versus* NRGO cathode in tap water.

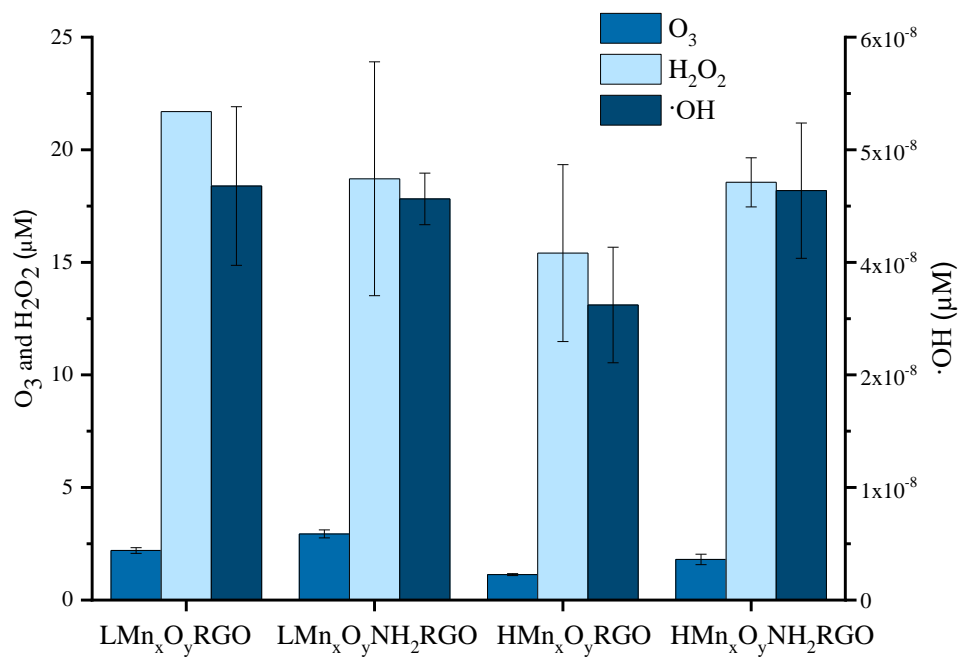

**Figure S11.** Concentrations of the electrochemically produced oxidants ($O_3$, $H_2O_2$ and ·OH) in steady state at 29 A m$^{-2}$ in tap water for $LMn_xO_yRGO$, $LMn_xO_yNH_2RGO$, $HMn_xO_yRGO$ and $HMn_xO_yNH_2RGO$ anode versus NRGO cathode.

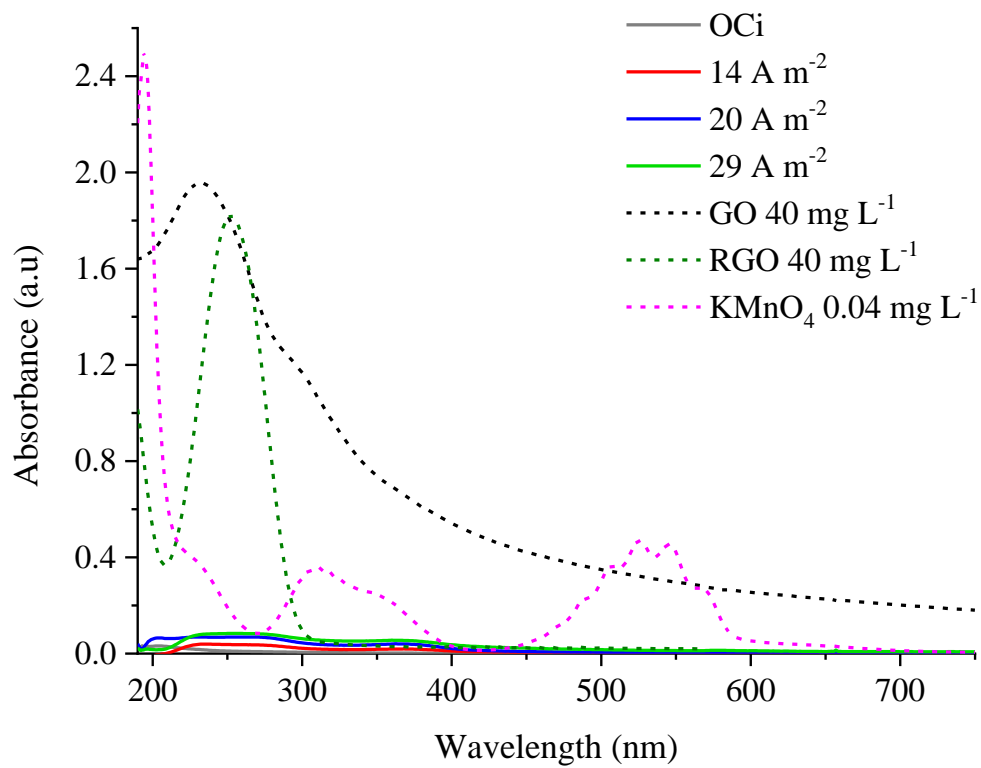

**Figure S12.** UV-Vis spectra of GO solution (40 mg L$^{-1}$), RGO dispersion (40 mg L$^{-1}$), KMnO$_4$ solution (0.04 mg L$^{-1}$) and the effluent of the reactor HMn$_x$O$_y$NH$_2$RGO at initial open circuit (OCi) and at 14, 20 and 29 A m$^{-2}$ using tap water as electrolyte.

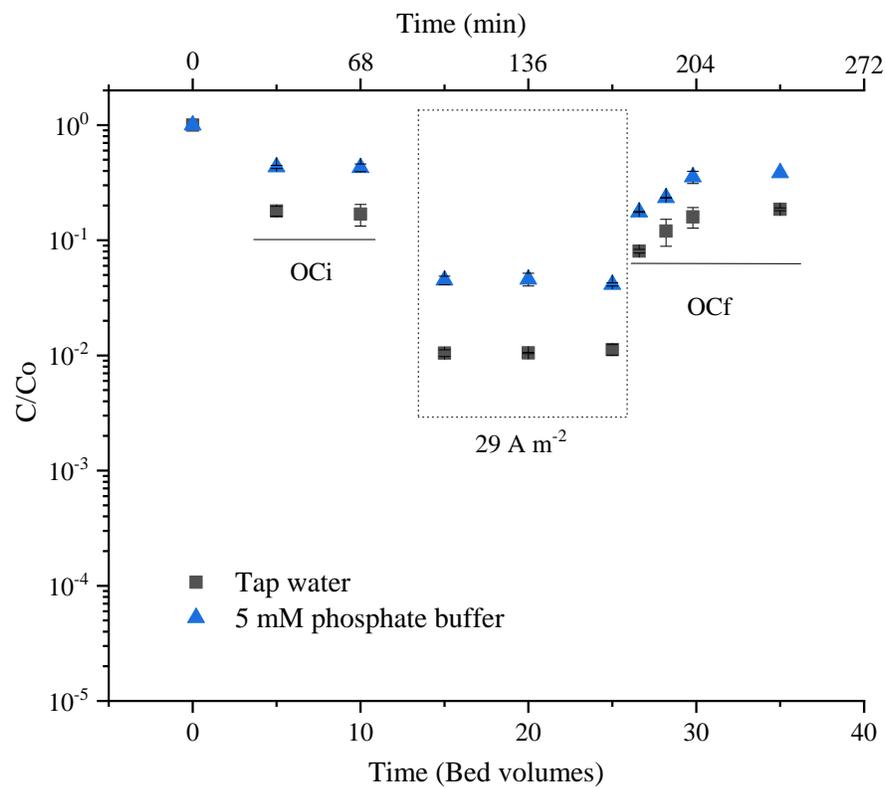

**Figure S13.** Electrochemical inactivation of *E. coli* ATCC 25922 in phosphate buffer (5 mM) and tap water, in the initial open circuit ($OC_i$), 29 A m$^{-2}$ and final OC ($OC_f$), using $HMn_xO_yNH_2RGO$ anode, versus NRGO cathode.

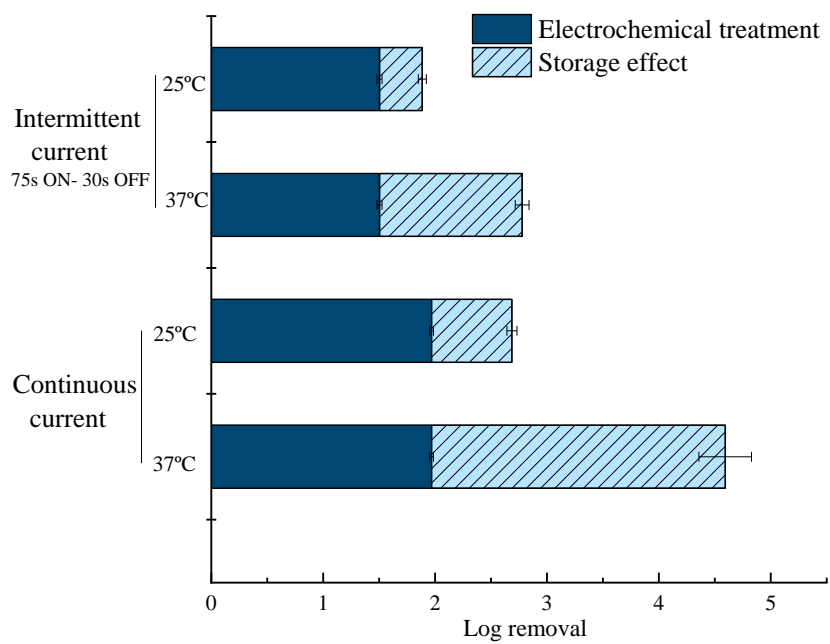

**Figure S14**. Inactivation of *E. coli* during the electrochemical treatment at 29 A m$^{-2}$ at continuous and intermittent current (75s ON-30s OFF) application in tap water using HMn$_x$O$_y$RGO as cathode and after 18h of storage at 25 and 37 °C.

**Table S1.** Literature examples on metal or metal oxide doped with amino groups attached to a carbon source and their applications.

| Compound | Application | Reference |
|---|---|---|
| $MnO_2$-$NH_2$-GO | Catalytic ozonation of an antibiotic (cephalexin) | [3] |
| $MnO_2$-$NH_2$/GO/p-$C_3N_4$ | Catalytic ozonation and photooxidation of cephalexin | [4] |
| $MnO_2$ nanowires on amine functionalized graphite | Catalysts for organic oxidation reactions | [5] |
| UiO-66-$NH_2$/RGO | Photocatalytic reduction of $CO_2$ | [6] |
| Fe@N-doped porous carbon | Oxidation of organic contaminants | [7] |

**Table S2.** Characteristics of the tap water used in the experiments.

| | |
|---|---|
| $PO_4^{3-}$ (mg $L^{-1}$) | < 0.008 |
| $Cl^-$ (mg $L^{-1}$) | 36.3 |
| $Na^+$ (mg $L^{-1}$) | 24 |
| $Ca^{2+}$ (mg $L^{-1}$) | 52.6 |
| Total organic carbon (TOC) (mg $L^{-1}$) | 3.7 |
| Conductivity (mS $cm^{-1}$) | 0.42 |
| pH | 7.6 |

**Table S3.** X-ray photoelectron spectroscopy (XPS) peak analysis of $Mn_xO_y$, $Mn_xO_yNH_2$ and the four-manganese oxide-doped graphene sponges (i.e., $LMn_xO_yRGO$, $LMn_xO_yNH_2RGO$, $HMn_xO_yRGO$ and $HMn_xO_yNH_2RGO$), including the binding energies of the deconvoluted peaks of Mn 3s, the peak splitting values, the average oxidation state and the oxidation states of the Mn species. The average oxidation state was calculated in **Figure S3**.

| Sponges | E1 (eV) | E2 (eV) | ΔE (eV) | Average $n^e$ | $n^e$ |
|---|---|---|---|---|---|
| $Mn_xO_y$ | 89.0 | 84.2 | 4.8 | 4.0 | 4 |
| $Mn_xO_yNH_2$ | 89.4 | 85.0 | 4.4 | 4.7 | 4 |
| $LMn_xO_yRGO$ | 88.2 | 83.3 | 4.8 | 4.0 | 4 |
| $LMn_xO_yNH_2RGO$ | 89.5 | 85.1 | 4.4 | 4.8 | 4 |
| $HMn_xO_yRGO$ | 89.7 | 84.3 | 5.5 | 2.6 | 2, 3 |
| $HMn_xO_yNH_2RGO$ | 89.6 | 84.1 | 5.5 | 2.6 | 2, 3 |

**Table S4.** Elemental composition (% of C, O, N and Mn) of the GO solution (data taken from Baptista-Pires et al[8]), $Mn_xO_y$, $Mn_xO_yNH_2$ and the anodic sponges ($LMn_xO_yRGO$, $LMn_xO_yNH_2RGO$, $HMn_xO_yRGO$ and $HMn_xO_yNH_2RGO$) by X-ray photoelectron spectroscopy (XPS).

|  | % C | % O | % N | % Mn |
|---|---|---|---|---|
| GO solution[8] | 62.6 | 36.9 | 0.9 | 0 |
| $Mn_xO_y$ | 0 | 49.9 | 2.1 | 48 |
| $Mn_xO_yNH_2$ | 22.5 | 38.7 | 5.1 | 33.7 |
| $LMn_xO_yRGO$ | 71.1 | 22.4 | 0.5 | 6.0 |
| $LMn_xO_yNH_2RGO$ | 52 | 34 | 2.9 | 11.1 |
| $HMn_xO_yRGO$ | 17.2 | 53.5 | 0.7 | 28.6 |
| $HMn_xO_yNH_2RGO$ | 34.1 | 45.2 | 2.8 | 17.9 |

**Table S5.** Intensity of the oxygen species obtained from the deconvoluted O 1s spectra of XPS of $Mn_xO_y$, $Mn_xO_yNH_2$ and the four anodic sponges ($LMn_xO_yRGO$, $LMn_xO_yNH_2RGO$, $HMn_xO_yRGO$ and $HMn_xO_yNH_2RGO$).

|  | % Mn-O-Mn | % defective O | % C-O/C=O | % adsorbed $H_2O$ |
|---|---|---|---|---|
| $Mn_xO_y$ | 50.6 | 38.1 | 0 | 11.3 |
| $Mn_xO_yNH_2$ | 25.5 | 31.3 | 30.1 | 13.1 |
| $LMn_xO_yRGO$ | 11.9 | 21.0 | 32.4 | 34.7 |
| $LMn_xO_yNH_2RGO$ | 7.5 | 23.3 | 31.8 | 37.4 |
| $HMn_xO_yRGO$ | 11.4 | 52.4 | 23.5 | 12.7 |
| $HMn_xO_yNH_2RGO$ | 8.0 | 36.7 | 31.0 | 24.3 |

**Table S6.** Ohmic drop-corrected anode potential ($V_{an}$) and the total cell potential ($V_{cell}$) of RGO, $LMn_xO_yRGO$ and $HMn_xO_yRGO$ at 58, 115 and 173 A m$^{-2}$ in 10 mM phosphate buffer.

|  | 58 A m$^{-2}$ | | 115 A m$^{-2}$ | | 173 A m$^{-2}$ | |
|---|---|---|---|---|---|---|
|  | $V_{an}$ (V/SHE) | $V_{cell}$ (V) | $V_{an}$ (V/SHE) | $V_{cell}$ (V) | $V_{an}$ (V/SHE) | $V_{cell}$ (V) |
| RGO | 2.1 | 7.5 | 2.2 | 9.8 | 2.3 | 12.2 |
| $LMn_xO_yRGO$ | 2.1 | 9.7 | 2.2 | 11.7 | 2.4 | 12.4 |
| $HMn_xO_yRGO$ | 4.3 | 10.3 | 4.5 | 12.5 | 4.8 | 14.8 |

**Table S7.** Ohmic drop-corrected anode potential ($V_{an}$), the total cell potential ($V_{cell}$) and energy consumption (E) of the four systems studied at 29 A m$^{-2}$ in tap water.

|  | $V_{an}$ (V/SHE) | $V_{cell}$ (V) | E (kWh m$^{-3}$) |
|---|---|---|---|
| LMn$_x$O$_y$RGO | 1.83 | 8.33 | 1.39 |
| LMn$_x$O$_y$NH$_2$RGO | 1.67 | 7.17 | 1.20 |
| HMn$_x$O$_y$RGO | 2.71 | 9.7 | 1.62 |
| HMn$_x$O$_y$NH$_2$RGO | 1.70 | 7.5 | 1.25 |

# References

[1] P. Hosseini-Benhangi, C.H. Kung, A. Alfantazi, E.L. Gyenge, Controlling the Interfacial Environment in the Electrosynthesis of MnOx Nanostructures for High-Performance Oxygen Reduction/Evolution Electrocatalysis, ACS Appl. Mater. Interfaces. 9 (2017) 26771–26785. https://doi.org/10.1021/acsami.7b05501.

[2] E. Cuervo Lumbaque, L. Baptista-Pires, J. Radjenovic, Functionalization of graphene sponge electrodes with two-dimensional materials for tailored electrocatalytic activity towards specific contaminants of emerging concern, Chem. Eng. J. 446 (2022). https://doi.org/10.1016/j.cej.2022.137057.

[3] J. Xu, Y. Li, M. Qian, J. Pan, J. Ding, B. Guan, Amino-functionalized synthesis of MnO2-NH2-GO for catalytic ozonation of cephalexin, Appl. Catal. B Environ. 256 (2019). https://doi.org/10.1016/j.apcatb.2019.117797.

[4] B. Yang, B. Guan, Synergistic catalysis of ozonation and photooxidation by sandwich structured MnO2-NH2/GO/p-C3N4 on cephalexin degradation, J. Hazard. Mater. 439 (2022) 129540. https://doi.org/10.1016/j.jhazmat.2022.129540.

[5] A. Chakravarty, D. Sengupta, B. Basu, A. Mukherjee, G. De, MnO2 nanowires anchored on amine functionalized graphite nanosheets: Highly active and reusable catalyst for organic oxidation reactions, RSC Adv. 5 (2015) 92585–92595. https://doi.org/10.1039/c5ra17777d.

[6] X. Zhao, M. Xu, X. Song, W. Zhou, X. Liu, Y. Yan, P. Huo, Charge separation and transfer activated by covalent bond in UiO-66-NH2/RGO heterostructure for CO2 photoreduction, Chem. Eng. J. 437 (2022). https://doi.org/10.1016/j.cej.2022.135210.

[7] C. Liu, Y. Wang, Y. Zhang, R. Li, W. Meng, Z. Song, F. Qi, B. Xu, W. Chu, D. Yuan, B. Yu, Enhancement of Fe@porous carbon to be an efficient mediator for peroxymonosulfate activation for oxidation of organic contaminants: Incorporation NH2-group into structure of its MOF precursor, Chem. Eng. J. 354 (2018) 835–848. https://doi.org/10.1016/j.cej.2018.08.060.

[8] L. Baptista-Pires, G.F. Norra, J. Radjenovic, Graphene-based sponges for electrochemical degradation of persistent organic contaminants, Water Res. 203 (2021). https://doi.org/10.1016/j.watres.2021.117492.